\documentclass[twocolumn,showpacs,preprintnumbers,
amsmath,amssymb,aps,prb,superscriptaddress]{revtex4}
\usepackage{graphicx}
\usepackage[latin1]{inputenc}

\usepackage{epstopdf,color}

\definecolor{red}{rgb}{1.,0.,0.}

\newcommand{\XXX}[1]{}
\newcommand{\hc}{\mbox{h.c.}}
\newcommand{\dagg}{^{\vphantom{\dagger}}}

\begin{document}
\title{Competing orders in the generalized Hund chain model
at half-filling}
\author{H. Nonne}
\affiliation{Laboratoire de Physique Th\'eorique et
Mod\'elisation, CNRS UMR 8089,
Universit\'e de Cergy-Pontoise, Site de Saint-Martin,
F-95300 Cergy-Pontoise Cedex, France}
\author{E. Boulat}
\affiliation{
Laboratoire Mat\'eriaux et Ph\'enom\`enes Quantiques,
Universit\'e Paris  Diderot, 2 Place Jussieu,
75205 Paris Cedex 13, France}
\author{S.\ Capponi}
\affiliation{Laboratoire de Physique Th{\'e}orique, Universit{\'e} de Toulouse,
UPS (IRSAMC), F-31062 Toulouse, France}
\affiliation{CNRS, LPT (IRSAMC), F-31062 Toulouse, France}
\author{P. Lecheminant}
\affiliation{Laboratoire de Physique Th\'eorique et
Mod\'elisation, CNRS UMR 8089,
Universit\'e de Cergy-Pontoise, Site de Saint-Martin,
F-95300 Cergy-Pontoise Cedex, France}

\date{\today}
\pacs{
71.10.Pm ; 71.10.Fd
}

\begin{abstract}
By using a combination of several non-perturbative techniques -- a one-dimensional
field theoretical approach together with numerical simulations using density matrix
renormalization group -- we present an extensive study
of the phase diagram of the generalized Hund model at half-filling.
This model encloses the physics of various
strongly correlated one-dimensional systems, such as two-leg electronic ladders,
ultracold degenerate fermionic gases carrying a large hyperfine spin $\frac{3}{2}$,
other cold gases like Ytterbium 171 or alkaline-earth condensates.
A particular
emphasis is laid on the possibility to enumerate and exhaust the eight possible Mott
insulating phases by means of a duality approach. We exhibit a one-to-one correspondence between
 these phases and those of the two-leg electronic ladders with interchain hopping.
Our results obtained from a weak coupling analysis are  in
remarkable quantitative agreement with our numerical results carried out at moderate coupling.
\end{abstract}

\maketitle
\section{Introduction}

A major focus of the study of strongly correlated
electronic systems is the analysis of the competition
between qualitatively distinct ground states and
the associated quantum phase transitions (QPT)~\cite{sachdev} in low dimensions.
A reason to concentrate on these matters stems from the hope that the
criticality of the system at such QPT's possibly results
in a universal description of their vicinity.
In $2+1$ space-time dimensions,
the resulting relativistic quantum field theories which describe
the zero-temperature transition between these quantum phases
are in general strongly coupled and can be highly non-trivial~\cite{senthil}.

In one dimension, the situation is much simpler
since the quantum critical points in standard condensed matter
systems are characterized by conformal field theories (CFT), which
often admit a simple free-field representation in terms
of free bosons or fermions.
In this respect, the bosonization approach has
been very successful to investigate
the physical properties of one-dimensional quantum phases~\cite{bookboso,giamarchi}.
Within this approach, several conventional and
exotic long-range ordered phases
have been revealed over the years
in two-leg ladder models~\cite{Linso8,szh,frahm,lehur,marston1,marston2,furusaki,wu04,momoi,assarafdemirep,
fabrizio,Lee,marston3,bunder}
and carbon nanotube systems~\cite{linCarbone,nersesyan03,
nersesyan07,bunder08} at half-filling.
Two different classes of Mott-insulating phases have
been found at half-filling in these systems.
A first class is two-fold degenerate corresponding
to the spontaneous breaking of discrete symmetries
such as the translation symmetry in charge density wave (CDW)
and bond-ordering phases, or the time-reversal symmetry
in d-density wave (DDW) phase.
In contrast,
the second class of Mott-insulating phases is non-degenerate.
A paradigmatic example of this class is the Haldane phase of the spin-1
Heisenberg chain~\cite{haldane} and of the two-leg spin ladder which breaks spontaneously a
non-local Z$_2$ $\times$ Z$_2$ discrete symmetry~\cite{tasaki,shelton,string2leg}.

Another striking particularity of many one-dimensional electronic systems
is the existence of hidden duality symmetries, within the
low-energy approach, which relate
many of the competing orders to a conventional
one like the CDW~\cite{Linso8,schulz,momoi,lee,controzzi,totsuka,boulat}.
Actually,
a general duality approach has been introduced recently
to describe the zero-temperature spin-gapped phases
of one-dimensional (1D) electronic systems away from half-filling~\cite{boulat}.
In this paper, we apply this approach to half-filled
systems of four-component fermions and revisit the problem
of competing orders in half-filled two-leg electronic ladders.
In this particular case, some of the duality symmetries
already exist at the level of the lattice model, and have been
first revealed in Ref.~\onlinecite{momoi}.
In addition to these, as it will be seen,
there are also interesting \emph{emergent duality symmetries}
which relate non-degenerate Mott insulating phases
to conventional order such as CDW. Unlike the former ones, those dualities do not bear a local
representation on the lattice.

The starting point of the duality approach to competing orders is to
identify the internal global symmetry group H of the lattice
model. For two-leg electronic ladders, or more generally
two-band models, the building blocks of the model are four-component
fermionic creation operators on each site $i$:
$c^{\dag}_{l\sigma,i}$ where $l=1,2$ is the
leg or orbital index  and $\sigma = \uparrow, \downarrow$
denotes the spin-$\frac{1}{2}$ index.
Three basic global continuous symmetries are retained:
a U(1) charge symmetry ($c_{l\sigma,i} \rightarrow e^{ i \phi} c_{l\sigma,i}$),
a SU(2) spin-rotational invariance ($c_{l\sigma,i} \rightarrow
\sum_{\sigma^{'}} (e^{i {\vec \theta} \cdot {\vec \sigma}/2})_{\sigma\sigma^{'}}
c_{l\sigma^{'},i}$, ${\vec \sigma}$ being
the Pauli matrices), and a U(1) orbital symmetry
($c_{1(2)\sigma,i} \rightarrow e^{ \pm i \phi} c_{1(2)\sigma,i}$).
Moreover, we will consider models for which the two legs, or two bands
behave identically; in other words, we impose a Z$_2$ invariance
under the permutation of the legs.
If we restrict ourselves to on-site interactions, the most general model
with H $=$ U(1)$_{c}$ $\times$ SU(2)$_{s}$ $\times$
U(1)$_{o}$ $\times$ Z$_2$ invariance then reads as follows~\cite{arovas}:
\begin{eqnarray}
  \mathcal{H}&=&-t \sum_{i,l\sigma}
  \left(c^\dag_{l\sigma,i}
  c^{\phantom\dag}_{l\sigma,i+1}
  + \hc \right)
  -\mu \sum_{i} n_i
  +\frac{U}{2}\sum_{i} n_i^2 \nonumber\\
  &+& J_H \sum_i \vec{S}_{1,i}
  \cdot\vec{S}_{2,i} + J_t \sum_i (T_i^z)^2,
  \label{GHundmodellattice}
\end{eqnarray}
with $n_i = \sum_{l\sigma} n_{l \sigma,i}$
($n_{l \sigma,i} = c^{\dag}_{l \sigma,i}
c_{l \sigma,i}\dagg$) being the occupation number on the $i^{\mbox{\scriptsize th}}$
site. In Eq.~(\ref{GHundmodellattice}), the spin operator
on leg $l$ is defined by
\begin{equation}
\vec{S}_{l,i}= \frac{1}{2} \sum_{\alpha, \beta} c_{l\alpha,i}^\dag
\vec{\sigma}_{\alpha\beta}
c^{\phantom\dag}_{l\beta,i},
\label{spinop}
\end{equation}
whereas $T_i^z=\frac{1}{2}\sum_{\sigma}
(n^{\phantom\dag}_{1\sigma,i}
-n^{\phantom\dag}_{2\sigma,i})$ is the generator
of the U(1) symmetry for orbital degrees of freedom.

Model (\ref{GHundmodellattice}) depends on three
microscopic couplings: a
Coulombic interaction $U$, a Hund coupling $J_H$,
and an ``orbital crystal field anisotropy'' $J_t$.
When $J_t =0$, the resulting model is the so-called Hund
model which has been investigated in the context
of orbital degeneracy~\cite{arovas,lee,fabrizio}.
The generalized Hund model (\ref{GHundmodellattice})
is directly linked to ultracold fermionic $^{171}$Yb and alkaline-earth
atoms with nuclear spin $I=\frac{1}{2}$~\cite{fukuhara,gorshkov}.
The two-orbital states are described in these systems by the ground
state ($^{1}$S$_{0}$ $\equiv$ g) and a long-lived excited state
($^{3}$P$_{0}$ $\equiv$ e).
The almost perfect decoupling of the
nuclear spin from the electronic angular momentum $J$ in the
two $e,g$ states ($J=0$ states) makes the s-wave scattering lengths of the problem
independent of the nuclear spin.
The low-energy Hamiltonian relevant to the $^{171}$Yb cold gas
loaded into a 1D optical lattice
 then reads~\cite{gorshkov}:
\begin{eqnarray}
{\cal H}_{\rm Yb} &=&-t \sum_{i,l\sigma}
  \left(c^\dag_{l\sigma,i}
  c^{\phantom\dag}_{l\sigma,i+1}
  + \hc \right)
+ \frac{{\tilde U}}{2}
\sum_{i,l} n_{l,i} ( n_{l,i} - 1)
\nonumber \\
&+& V \sum_{i} n_{g,i} n_{e,i} +
V_{ex} \sum_{i,\alpha,\beta}
c^{\dagger}_{g \alpha,i} c^{\dagger}_{e \beta,i}
c_{g \beta,i}\dagg c_{e \alpha,i}\dagg,
\label{ytterbium}
\end{eqnarray}
where $c^{\dagger}_{l \sigma,i}$ is the fermionic
creation operator at site $i$ with the nuclear spin-$\frac{1}{2}$ index $\sigma = \uparrow,
\downarrow$ in the $l=e,g$ electronic states.
The occupation number of electronic states is $n_{l,i} = \sum_{\sigma}  c^{\dag}_{l \sigma,i}
c_{l \sigma,i}\dagg$.
Model (\ref{ytterbium}) is then directly equivalent
to the generalized Hund model (\ref{GHundmodellattice})
with the correspondence:
${\tilde U} = U + J_t/2$, $V = U - J_t/2 - J_H/4$, and $V_{ex} = -J_H/2$.

Apart from this connection to cold fermions physics,
one of the main interests of model (\ref{GHundmodellattice})
stems from the fact that it contains a large
variety of relevant models with extended continuous symmetries,
some of which having appeared in different contexts.
First of all, when $J_t= - 3 J_H/4$, the continuous symmetry is
promoted to U(1)$_{c}$
$\times$ SU(2)$_{s}$ $\times$ SU(2)$_{o}$,
and one recovers the so-called spin-orbital
model~\cite{yamashita,khomskii,mila,aza4spin,itoi00}.
In absence of Hund coupling, i.e., $J_H=0$,
the continuous symmetry group of model (\ref{GHundmodellattice})
is U(1)$_{c}$ $\times$  U(1)$_{o}$
$\times$ SO(4)$_{s}$ where each chain has a separate
SU(2) spin-rotational symmetry.
When $J_t=J_H/4$, model (\ref{GHundmodellattice}) displays
an SO(5) symmetry which unifies spin and orbital
degrees of freedom. The resulting model, with
U(1)$_{c}$ $\times$ SO(5)$_{s,o}$ continuous symmetry,
 is relevant to four-component fermionic
cold atom systems~\cite{zhang,Lecheminant2005,Wu2005,tsvelik06,tu,wu06,sylvainquart,capponiMS,roux,jiang,heloise}.
Finally, when $J_H=J_t=0$, one recovers the U(4) Hubbard model,
that has been extensively analyzed in recent years~\cite{assarafdemirep,assaraf,solyom,ueda}.
As it will be seen in Section II, at half-filling, many
other highly symmetric lines can be identified.
For instance, the line $J_H= 8 U$ unifies spin and charge degrees of freedom
with an extended U(1)$_{o}$ $\times$ SO(5)$_{s,c}$ continuous symmetry~\cite{sylvainzhang}.
The corresponding model has been previously  introduced
by Scalapino, Zhang and Hanke (SZH)~\cite{szh} in connection to the SO(5) theory
which relates antiferromagnetism to d-wave
superconductivity~\cite{zhangso5}.

In this paper, we will investigate the nature of the insulating
phases of the zero-temperature phase diagram of
model (\ref{GHundmodellattice}) at half-filling.
In this respect,
it will be shown that the duality approach
of Ref.~\onlinecite{boulat}
for half-filled fermions with internal global
symmetry group H = U(1)$_{c}$ $\times$
SU(2)$_{s}$ $\times$ U(1)$_{o}$ $\times$ Z$_2$
yields eight fully gapped phases.
These eight Mott-insulating phases fall into
two different classes.
On the one hand, the first class consists of four doubly degenerate phases
which spontaneously break a discrete symmetry of the underlying lattice model.
On the other hand, the second class contains four non-degenerate Mott insulating
phases. A first one is the rung-singlet (RS) phase where two spins on each rung
lock into a singlet. A second non-degenerate phase is a rung-triplet (RT) phase
where the spins now combine into a triplet and this phase is known
to be adiabatically connected to the Haldane phase of the spin-1 Heisenberg
chain~\cite{string2leg}. Finally, two other
Haldane-like phases are found: they are spin-singlet
but involve two different pseudo-spin 1 operators which are respectively built from
charge and orbital degrees of freedom.
In the case of the charge pseudo-spin 1 operator,
the resulting Haldane charge (HC) phase has been found very recently in
the context of 1D half-filled spin-$\frac{3}{2}$
cold fermions~\cite{heloise}.

In addition to this duality approach, it will be shown, by means of a
one-loop renormalization group (RG) approach and numerical simulations
(using the Density Matrix Renormalization Group (DMRG)
algorithm~\cite{DMRG}), that the zero-temperature phase diagram of
model (\ref{GHundmodellattice}) displays seven out of the eight
expected insulating phases.  We find it remarkable that model
(\ref{GHundmodellattice}),
that only has three
independent coupling constants, turns out to have a rich phase diagram
which includes the four non-degenerate Mott insulating phases.

Finally, we will make contact with the eight Mott-insulating phases
found over the years in half-filled generalized two-leg ladders with a
$t_{\perp}$ transverse hopping term~\cite{furusaki,wu04}.  The latter
term breaks explicitly the U(1)$_{o}$ symmetry but it is known that
this symmetry is recovered at low-energy~\cite{Linso8,wu04}. The
relevant global symmetry group is still H and the same duality
approach thus applies to that case.  In this respect, we will connect
the two families of eight fully gapped phases found for $t_{\perp}=0$
and for $t_{\perp}\neq0$.  In particular, it will be shown that
the two problems are in fact connected by an emergent non-local
duality symmetry.

The rest of the paper is organized as follows.
In Section II, we discuss the symmetries of
model (\ref{GHundmodellattice}).
We also present  a strong-coupling analysis along
special highly-symmetric lines which gives some clues
about the nature of the non-degenerate Mott-insulating phases.
The low-energy investigation is then presented in Section III.
It contains the duality approach to half-filled fermions
with internal symmetry group H $=$ U(1)$_{c}$ $\times$
SU(2)$_{s}$ $\times$ U(1)$_{o}$ $\times$ Z$_2$.
The zero-temperature phase diagram of
the generalized Hund model (\ref{GHundmodellattice})
and that of highly symmetric models are deduced
by a one-loop RG analysis.
We then connect our results to the known insulating phases
of generalized two-leg ladder models with a  $t_{\perp}$ hopping term.
In Section IV, we map out the phase diagram of model (\ref{GHundmodellattice}),
SZH, and spin-orbital models with $t_{\perp}=0$
by means of DMRG calculations to complement the low-energy approach.
Our concluding remarks are presented in Section V.
The paper is supplied with three appendices which provide
some additional information.
Appendix A describes the technical details of the continuum limit
of model (\ref{GHundmodellattice}). The low-energy approach of edge states
in the non-degenerate Mott-insulating phases are discussed in
Appendix B. Finally, Appendix C presents the main effect of the interchain
hopping in the strong-coupling regime, close to the orbital symmetric line.

\section{Symmetries and strong coupling} \label{sym}

Before investigating the zero-temperature phase diagram of the
generalized Hund model
by means of the low-energy and DMRG approaches, it is important
to fully determine the special lines which exhibit enlarged symmetry.
It turns out that, at half-filling, many highly symmetric lines
can be highlighted.
Their study will give some important
clues on the possible Mott-insulating phases of
the model at half-filling.

\subsection{Highly symmetric lines}

The generalized Hund model (\ref{GHundmodellattice})
enjoys a global internal symmetry group H $=$ U(1)$_{c}$ $\times$
SU(2)$_{s}$ $\times$ U(1)$_{o}$ $\times$ Z$_2$
on top of the lattice discrete symmetries like one-step
translation invariance, time-reversal symmetry, site and link
parities.
For generic filling, and on four different manifolds
-- that correspond to some fine-tuning of the lattice couplings --
in the space of coupling constants,
this model possesses a higher symmetry.

First of all, on top of the SU(2)$_s$ that rotates
spin degrees of freedom,
one can define a SU(2)$_o$ orbital pseudo-spin
operator:
\begin{eqnarray}
  T_i^\dag&=&c_{1\uparrow,i}^\dag
  c_{2\uparrow,i}\dagg +c_{1\downarrow,i}^\dag
  c_{2\downarrow,i}\dagg\nonumber\\
  T_i^z&=&\frac{1}{2}\left(n_{1,i}
  -n_{2,i}\right),
  \label{orbitalpseudospin}
\end{eqnarray}
with $n_{l,i} = \sum_{\sigma} n_{l \sigma,i}$, $l = 1,2$.
When $J_t = - 3 J_H/4$, the U(1) orbital symmetry of the Hund model
(\ref{GHundmodellattice})
is enlarged to SU(2)$_o$,
with generators given by Eq. (\ref{orbitalpseudospin}).
The resulting model displays a U(1)$_c$ $\times$ SU(2)$_s$ $\times$ SU(2)$_o$
continuous symmetry and has been considered in systems with orbital
degeneracy like the spin-orbital model~\cite{arovas,yamashita,khomskii,mila,aza4spin,itoi00} .

A second highly symmetric model is defined for $J_H = 0$: then
the interacting part of model (\ref{GHundmodellattice}) simplifies
as follow:
\begin{eqnarray}
{\cal H}^{\rm int}_{SO(4)} &=& \frac{1}{2} \left(U + \frac{J_t}{2} \right)
\sum_i \left( n_{1,i}^2 + n_{2,i}^2 \right) \nonumber \\
&+& \left(U - \frac{J_t}{2} \right) \sum_i n_{1,i} n_{2,i},
\label{SO4model}
\end{eqnarray}
from which we deduce that each leg has a separate SU(2) spin rotation
symmetry so that the continuous symmetry group of model (\ref{SO4model}) is
U(1)$_c$ $\times$ U(1)$_o$ $\times$ SO(4)$_s$.

When $J_t = J_H/4$, as shown in Ref.~\onlinecite{sylvainzhang},
model (\ref{GHundmodellattice}) is known to be equivalent to the
spin-$\frac{3}{2}$ cold fermionic model with interacting part:
\begin{eqnarray}
{\cal H}^{\rm int}_{{\rm spin-}\frac{3}{2}} = U_0 \sum_i P^{\dagger}_{00,i}
P_{00,i}\dagg
+ U_2 \sum_i \sum_{m=-2}^{2} P^{\dagger}_{2m,i} P_{2m,i}\dagg,
\label{coldatoms}
\end{eqnarray}
where $U_0 = (2 U - 7 J_t)/4$, and $U_2 = (2 U + J_t)/4$.
In Eq.~(\ref{coldatoms}), we have $P^{\dagger}_{Jm,i} = \sum_{\alpha
\beta} \langle{J m|\frac{3}{2},\frac{3}{2};\alpha \beta}\rangle c^{\dagger}_{\alpha,i}
c^{\dagger}_{\beta,i}$, $\alpha,\beta = \pm \frac{3}{2}, \pm \frac{1}{2}$, and
$ \langle{J m|\frac{3}{2},\frac{3}{2};\alpha \beta}\rangle$ are the
Clebsch-Gordan coefficients for spin $\frac{3}{2}$.
Model (\ref{coldatoms}) is known to
exhibit a U(1)$_c$ $\times$
SO(5)$_{s,o}$ continuous symmetry without any fine-tuning~\cite{zhang}.

Finally, for $J_H=J_t =0$, spin and orbital
degrees of freedom unify to a maximal SU(4) symmetry and model (\ref{GHundmodellattice})
takes the form of the Hubbard model for four component
fermions with a U(4) invariance.

At half-filling, the chemical potential $\mu$
is given by $\mu_0=\frac{3U}{2}$ to ensure
particle-hole symmetry.
More highly-symmetric lines can be found
in this particle-hole symmetric case.
It stems from the fact that, as in the spin-$\frac{1}{2}$ Hubbard model,
 the U(1)$_c$ charge symmetry can be enlarged to
an SU(2)$_c$ symmetry at half-filling~\cite{yang,yangzhang}.
In this respect, one can define
a charge pseudo-spin operator by:
\begin{eqnarray}
  J_i^\dag&=&c_{1\uparrow,i}^\dag
  c_{2\downarrow,i}^{\dag}
  -c_{1\downarrow,i}^\dag
  c_{2\uparrow,i}^{\dag}\nonumber\\
  J_i^z&=&\frac{1}{2}\left(n_i - 2\right),
\label{chargesu2}
\end{eqnarray}
which is a SU(2)$_s$ spin-singlet that satisfies the SU(2) commutation relations.
This operator is the generalization in two-leg ladder or two-band systems
of the pseudo-spin $\frac{1}{2}$ operator introduced by Anderson\cite{anderson}
and by Yang in eta-pairing problems~\cite{yang}.

\begin{figure}[h]
\centering
\includegraphics[scale=0.25]{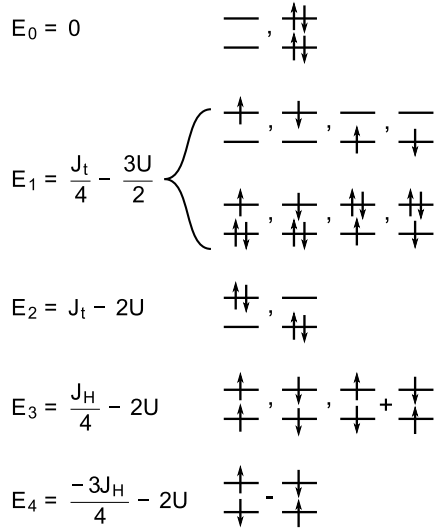}
\caption{Energy level diagram for the one-site
Hamiltonian (\ref{GHundmodellattice}) with $t=0$.}\label{energylevels}
\end{figure}

Many interesting lines can then be considered.
A simple way to reveal them is to consider the energy
levels of the one-site Hamiltonian (\ref{GHundmodellattice}) with
$t=0$. The corresponding spectrum is depicted
in Fig. \ref{energylevels}. On top of the four symmetric lines that we have identified above,
we find nine additional lines where higher continuous
symmetries emerge (see Table 1).
Among all these new highly symmetric lines,
there are three interesting models with two independent
coupling constants, i.e., with only one fine-tuning.

\begin{table*}
\caption{\label{extendedsymmetry}Extended continuous symmetries
of model (\ref{GHundmodellattice})
at half-filling}
\begin{ruledtabular}
\begin{tabular}{lll}
Extended continuous symmetry & Fine-tuning &Degenerate levels\\
\hline
$U(1)_c \times SU(2)_o\times SU(2)_s$ &
$J_t=-3J_H/4$ & $E_2=E_4$\\
$U(1)_c \times U(1)_o \times SO(4)_s$ &
$J_H=0$ & $E_3=E_4$\\
$U(1)_c \times SO(5)_{s,\,o}$ &
$J_t=J_H/4$ & $E_2=E_3$\\
$U(1)_c \times SU(4)_{s,\,o}$ &
$J_H=J_t=0$ & $E_2=E_3=E_4$\\
$SO(7)$ &
$J_t=2U,\, J_H=8U$ & $E_0=E_2=E_3$\\
$U(1)_o \times SU(4)_{s,\,c}$ &
$U=J_H=0$ & $E_0=E_3=E_4$\\
$SU(2)_c \times SO(5)_{s,\,o}$ &
$J_t=-2U/3,\, J_H=-8U/3$ & $E_0=E_4,\, E_2=E_3$\\
$SU(2)_o \times SO(5)_{s,\,c}$ &
$J_t=-6U,\, J_H=8U$ & $E_0=E_3,\, E_2=E_4$\\
$SO(5)_{c,\,o} \times SU(2)_s$ &
$J_t=2U,\, J_H=-8U/3$ & $E_0=E_2=E_4$\\
$U(1)_o \times SO(5)_{s,\,c}$ &
$J_H=8U$ & $E_0=E_3$\\
$SO(4)_{c,\,o} \times SO(4)_s$ &
$J_H=0,\, J_t=2U$ & $E_0=E_2,\, E_3=E_4$\\
$ SO(4)_{c,\,o} \times SU(2)_s$ &
$J_t=2U$ & $E_0=E_2$\\
$U(1)_o\times SU(2)_s \times SU(2)_c$&
$J_H=-8U/3$ & $E_0=E_4$\\
\end{tabular}
\end{ruledtabular}
\end{table*}

A first one corresponds to the SZH model with
U(1)$_o$ $\times$ SO(5) continuous symmetry.
Such SZH model with no transversal hopping $t_{\perp}$ is defined
as follows: $\mathcal{H}_{\rm SZH}=
\mathcal{H}_{t_\parallel}
+\mathcal{H}_{\textrm{rung}}$ with
\begin{eqnarray}
  &&\mathcal{H}_{t_\parallel}=
  -t
  \sum_{i,\,\sigma}(c_{\sigma,\,i}^\dag
  c_{\sigma,\,i+1}\dagg+d_{\sigma,\,i}^\dag
  d_{\sigma,\,i+1}\dagg+\hc), \nonumber\\
  &&\mathcal{H}_{\textrm{rung}}=U_{\textrm{SZH}}
  \sum_i \left((n^{\phantom\dag}_{c\uparrow,\,i}
  -\frac{1}{2})(n^{\phantom\dag}_{c\downarrow,\,i}
  -\frac{1}{2})+(c\rightarrow d)\right) \nonumber\\
  &&\qquad+V_{\textrm{SZH}}\sum_i
  (n^{\phantom\dag}_{c,\,i}-1)
  (n^{\phantom\dag}_{d,\,i}-1)\nonumber\\
  &&\qquad+J_{\textrm{SZH}}\sum_i
  \vec{S}_{c,\,i}\cdot \vec{S}_{d,\,i},
  \label{SZHmodel}
\end{eqnarray}
where $c_{\sigma}$ and $d_{\sigma}$ are respectively the fermion annihilation
operator of the upper
and lower leg of the ladder with spin index  $\sigma$.
The occupation numbers on the $i^{\mbox{\scriptsize th}}$ site are denoted by
$n_{c(d),i}$ respectively.
The spin operators ${\vec S}_{c(d),i}$ are defined similarly
to those of the Hund model (see Eq.~(\ref{spinop})).
It is straightforward to relate the SZH model
to the generalized Hund model (\ref{GHundmodellattice}):
\begin{eqnarray}
&&U=\frac{U_{SZH}+V_{SZH}}{2}
\nonumber\\
&& J_H=J_{SZH}
\nonumber\\
&& J_t=U_{SZH}-V_{SZH}.
\end{eqnarray}
As shown in Ref.~\onlinecite{szh}, the fine-tuning
$J_{\textrm{SZH}}=4(U_{\textrm{SZH}}
+V_{\textrm{SZH}})$ (i.e., $J_H=8U$
in the context of model
(\ref{GHundmodellattice})) makes the
lattice model (\ref{SZHmodel})
U(1)$_o$ $\times$ SO(5)$_{s,c}$ symmetric. The SO(5) symmetry unifies here
spin and charge degrees of freedom and is thus
different from the spin-$\frac{3}{2}$ cold fermionic atoms
(\ref{coldatoms}) case.

A second symmetric line is found
for $J_t = 2U$ with the emergence of a
SU(2)$_s$ $\times$ SO(4)$_{c,o}$ continuous symmetry.
In that case, charge and orbital degrees of freedom play a
symmetric role and are unified by a SO(4) symmetry.

Finally, the last extended symmetric ray with
the fine-tuning $J_H = - 8 U/3$ (see Table I) corresponds
to  a model with U(1)$_o$ $\times$
SU(2)$_c$ $\times$ SU(2)$_s$ continuous symmetry.
Such a model has two independent SU(2) symmetries:
one for the spin degrees of freedom and also a second for
the charge degrees of freedom.
In this respect, it is very similar to the
spin-orbital model and can be called ``charge-spin'' model.

\subsection{Strong-coupling analysis}
\label{strongcoupling}

The identification of these highly symmetric models
is very useful since several possible
insulating phases of the
generalized Hund model (\ref{GHundmodellattice})
can be inferred from a strong-coupling analysis.
Such an approach has already been performed for
some special lines of Table I such as
the half-filled U(4) Hubbard chain~\cite{affleckmarston,onufriev},
the SO(5) spin-$\frac{3}{2}$ model~\cite{wu06,heloise},
and the SZH one~\cite{szh,frahm}.

Here, we present a simple strong-coupling approach
along three special lines which enables us
to identify several non-degenerate Mott-insulating phases.
To this end, let us first consider the line
$J_t=2U$ with $U>0$ and $J_H<0$.
In the absence of hopping term (i.e., $t=0$),
the lowest energy states are the
spin triplet $E_3$ (see Fig. \ref{energylevels}).
An effective Hamiltonian can then be deduced by treating the
hopping term as a perturbation in  the strong coupling regime
$|U,J_H,J_t|\gg t$.
At second order
of perturbation theory, we find an
antiferromagnetic SU(2) Heisenberg chain:
\begin{equation}
  \mathcal{H}_{\text{eff}}=J_s \sum_{i} \left( \vec{S}_{1,i}
+ \vec{S}_{2,i} \right)
  \cdot \left( \vec{S}_{1,i+1}
+ \vec{S}_{2,i+1} \right),
\end{equation}
where $J_s=-4t^2/(J_H - 4 U) > 0$, and $\vec{S}_{l,i}$
are the spin operators defined in Eq.~(\ref{spinop}).
The resulting fully gapped phase is the well-known
RT phase of the two-leg spin-$\frac{1}{2}$ ladder
which is adiabatically connected to the Haldane
phase of the spin-1 chain~\cite{string2leg}.
Such a non-degenerate gapful phase
displays a hidden antiferromagnetic ordering
which is revealed by a string-order parameter~\cite{dennijs,shelton,kim}:
\begin{eqnarray}
\lim_{|i-j| \rightarrow \infty}
\langle \left(S^z_{1,i}
+ S^z_{2,i} \right)
e^{ i \pi \sum_{k=i+1}^{j-1}
\left(S^z_{1,k}
+ S^z_{2,k} \right)}
\nonumber \\
\times \left(S^z_{1,j}
+ S^z_{2,j} \right)
\rangle \ne 0 .
\label{stringRT}
\end{eqnarray}
This RT phase is also known to exhibit spin-$\frac{1}{2}$ edge states
when open-boundary conditions are considered~\cite{aklt,kennedy,orignac}.

A second interesting line is
$J_H=-8 U/3$ where the charge degrees of freedom
enjoy an SU(2) symmetry enlargement.
In the absence of hopping term,
the lowest energy states for $U<0$ and $J_t>0$
are the $E_0,E_4$ levels as it can be seen from Fig. \ref{energylevels}.
Keeping only these three states,
we obtain, at second order of perturbation theory,
an effective (pseudo) spin-1
antiferromagnetic SU(2) Heisenberg chain:
\begin{equation}
  \mathcal{H}_{\text{eff}}=J_c \sum_i \vec{J}_i
  \cdot \vec{J}_{i+1} ,
\label{heisenbergcharge}
\end{equation}
with $J_c=-4t^2/(6U-J_t) > 0$.
The effective Hamiltonian (\ref{heisenbergcharge})
expresses in terms of the spin-singlet charge operator
(\ref{chargesu2}) which is a pseudo spin-1 operator in the triplet
states $E_0,E_4$ of Fig. \ref{energylevels}.
We then expect the emergence of Haldane-like
phase for charge degrees of freedom
as it has been recently found in the context of
half-filled spin-$\frac{3}{2}$ cold fermions~\cite{heloise}.
Such a HC phase is fully gapped and non-degenerate.
It displays a hidden ordering that is revealed by
the string-order parameter:
\begin{equation}
\lim_{|i-j| \rightarrow \infty}
\langle J^z_{i}
e^{ i \pi \sum_{k=i+1}^{j-1}
J^z_k}
J^z_{j}
\rangle \ne 0.
\label{stringcharge}
\end{equation}
A deviation from the line $J_H=-8 U/3$ breaks the SU(2) charge symmetry
down to U(1) and in the strong-coupling regime the
lowest correction to model (\ref{heisenbergcharge})
is a single-ion anisotropy term
$D_c\sum_i (J_i^z)^2$ (with $D_c=3J_H/4+2U$).
The Haldane phase of the spin-1 chain is known to be stable under a weak single-ion
anisotropy~\cite{schulzspin1}. A large enough $D_c$ may give rise to a Ising
phase with $\langle J^z_i \rangle \ne 0$
(i.e. a CDW from the definition (\ref{chargesu2}))
or to a large-$D_c$ phase which is a non-degenerate
gapped singlet phase.
The latter, with $\langle J^z_i \rangle
= 0$, corresponds to the RS
phase of the two-leg spin-$\frac{1}{2}$ ladder where the two spins
of the rung bind into a singlet state ($E_4$ state
of Fig.~\ref{energylevels}) for an antiferromagnetic
interchain coupling.

Finally, a last interesting symmetric ray is
$J_t = - 3 J_H/4$ where the U(1) orbital symmetry
is enlarged to SU(2).
Along this line,
when $J_H>0$ and $U$ is not too
negative, the lowest energy states of the one-site Hamiltonian are
levels $E_2,E_4$ of Fig. \ref{energylevels}.
At second order of perturbation theory, we now find  a spin-1
antiferromagnetic SU(2) Heisenberg chain for the orbital
degrees of freedom:
\begin{equation}
  \mathcal{H}_{\text{eff}}=J_o
  \sum_i \vec{T}_i \cdot \vec{T}_{i+1},
\label{heisenbergorbital}
\end{equation}
with $J_o=16t^2/(9J_H+8U)$.
We thus expect the emergence of a new Haldane phase
for the orbital degrees of freedom that will be
called Haldane orbital (HO) phase in the rest of the paper.
The resulting hidden ordering is captured by the
following string-order parameter:
\begin{equation}
\lim_{|i-j| \rightarrow \infty}
\langle T^z_{i}
e^{ i \pi \sum_{k=i+1}^{j-1}
T^z_k}
T^z_{j}
\rangle \ne 0 .
\label{stringorbital}
\end{equation}
A deviation from
the line $J_t = - 3 J_H/4$ breaks the SU(2) orbital symmetry down to
U(1) and in the strong-coupling regime
the lowest correction to model (\ref{heisenbergorbital})
is a single-ion
anisotropy term $D_o\sum_i (T_i^z)^2$ (with
$D_o=J_t+3J_H/4$). For sufficiently strong
value of $|D_o|$, the HO phase will be
destabilized into either an orbital density wave (ODW)
which is described by the $E_2$ states with $\langle
T_i^z \rangle \ne 0$ or a RS phase, i.e., the $E_4$ state
with $\langle T_i^z \rangle = 0$.

In summary, the strong-coupling analysis along
highly symmetric lines reveals
the existence of four non-degenerate
Mott-insulating phases (RT, HC, HO, RS)
and two gapful phases with long-range density ordering (CDW, ODW).

\section{Low-energy Approach}

In this section, we present the details of the
low-energy approach of the generalized Hund
model (\ref{GHundmodellattice}) at half-filling,
in the weak-coupling regime $|U,J_H,J_t| \ll t$.
The zero-temperature phase diagram of model
(\ref{GHundmodellattice}) will be investigated
by means of the combination of a duality
approach and one-loop RG calculations.
In particular, we will determine the
different insulating phases in the weak-coupling
regime and make connection with the ones
found within the strong-coupling approach.

\subsection{Phenomenological approach}

The starting point of the low-energy approach is
the linearization around the Fermi points $\pm k_F$
of the dispersion relation for non-interacting
four-component fermions. We thus introduce four
left and right moving Dirac fermions
$L_{l\sigma}, R_{l\sigma}$ ($l=1,2$ and
$\sigma=\uparrow,\downarrow$), which describe
the lattice fermions $c_{l\sigma,i}$ in the
continuum limit:
\begin{equation}
  \frac{c^{\phantom\dag}_{l\sigma,i}}{\sqrt{a_0}}
  \to R_{l\sigma}(x) e^{ik_F x}+ L_{l\sigma}(x)
  e^{-i k_F x},\label{contfer}
\end{equation}
with $k_F = \pi/2a_0$ at half-filling and
$x= i a_0$ ($a_0$ being the lattice spacing).
The next step of the approach is to use
the Abelian bosonization of these Dirac fermions
to obtain the low-energy effective Hamiltonian
for model (\ref{GHundmodellattice}).
The details of these calculations are given in
Appendix A. Here, we present a more phenomenological
approach which is based on the symmetries of the lattice
model (\ref{GHundmodellattice}).

The continuous symmetry of the non-interacting model
(\ref{GHundmodellattice}) is SO(8). In the continuum limit,
the SO(8) symmetry can be revealed by introducing
eight real (Majorana) fermions from the four complex Dirac
$(R,L)_{l\sigma}$  ones.
The non-interacting fixed point is then described
by the SO(8)$_1$ CFT with central charge $c=4$~\cite{dms}.
The chiral currents $J_{L,R}^{(a,b)}$ ($1 \le a < b \le 8$),
which generates this CFT, can be expressed as fermionic
bilinears:
$J_{L(R)}^{(a,b)} = i
\xi^{a}_{L(R)} \xi^{b}_{L(R)}$, where $\xi^{a}_{L(R)}$
are the eight left (right) moving Majorana fermions.

When interactions are included, the SO(8) symmetry
is broken down to H $=$ U(1)$_{c}$ $\times$ SU(2)$_{s}$ $\times$
U(1)$_{o}$ $\times$ Z$_2$. The key point of the analysis is to identify
how the eight Majorana fermions of the SO(8)$_1$ CFT act in H.
One way to obtain the correspondence is to focus on the
currents which generate the different continuous symmetry
groups in H.  The uniform part of the continuum
limit of  the spin operator (\ref{spinop}) on the leg $l=1,2$
defines the chiral SU(2)$_1$ currents ${\vec J}_{lR,L}$:
\begin{equation}
{\vec J}_{l L} = L^{\dagger}_{l \alpha}
\frac{{\vec \sigma}_{\alpha \beta}}{2}
L_{l \beta}\dagg,
{\vec J}_{l R} = R^{\dagger}_{l \alpha}
\frac{{\vec \sigma}_{\alpha \beta}}{2}
R_{l \beta}\dagg.
\label{currentspinleg}
\end{equation}
As in two-leg spin ladder~\cite{shelton,allen}, the sum and difference
of these chiral SU(2)$_1$  currents can be locally expressed in terms
of four Majorana fermions $\xi^{1,2,3,6}_{R,L}$ among
the eight original ones:
\begin{eqnarray}
{\vec J}_{1 R,L} + {\vec J}_{2 R,L} &=&
- \frac{i}{2} \; {\vec \xi}_{R,L} \wedge {\vec \xi}_{R,L} \nonumber \\
{\vec J}_{1 R,L} - {\vec J}_{2 R,L} &=&
i  \; {\vec \xi}_{R,L} \xi^{6}_{R,L} ,
\label{su2curmajo}
\end{eqnarray}
where the triplet of Majorana fermions
${\vec \xi} = (\xi^1, \xi^2, \xi^3)$ accounts
for the spin degrees of freedom since
the SU(2)$_s$ spin rotation symmetry of
the lattice model (\ref{GHundmodellattice}) is generated in the continuum
by ${\vec J}_{1 R} + {\vec J}_{2 R} + R \rightarrow L$.
The  Majorana fermion $\xi^{6}$ is related to
the discrete Z$_2$ interchain exchange as it can be seen
from Eq.~(\ref{su2curmajo}).
Finally, the four remaining Majorana fermions
can be cast into two pairs, each of which is associated
to the two U(1) symmetries in H:
$\xi^{4,5}$ (respectively $\xi^{7,8}$) Majorana fermions describe the
orbital (respectively charge) U(1) symmetry.

With this identification at hand, we can derive
the low-energy effective theory for the
generalized Hund model (\ref{GHundmodellattice})
at half-filling. Assuming only four-fermion (marginal)
interactions, the most general model with
H $=$ U(1)$_{c}$ $\times$ SU(2)$_{s}$ $\times$
U(1)$_{o}$ $\times$ Z$_2$ invariance can be easily deduced
from the Majorana fermion formalism:
\begin{eqnarray}
  \mathcal{H}&=& - \frac{i v_c}{2}\sum_{a=7}^8
  (\xi_R^a \partial_x \xi_R^a
  - \xi_L^a \partial_x \xi_L^a)\nonumber\\
  &-&\frac{i v_s}{2}\sum_{a=1}^3 (\xi_R^a
  \partial_x \xi_R^a - \xi_L^a \partial_x
  \xi_L^a)\nonumber\\
  &-&\frac{i v_t}{2}\sum_{a=4}^5 (\xi_R^a
  \partial_x \xi_R^a - \xi_L^a \partial_x
  \xi_L^a)\nonumber\\
  &-&\frac{i v_0}{2} (\xi_R^6 \partial_x
  \xi_R^6 - \xi_L^6 \partial_x \xi_L^6)\nonumber\\
  &+&\frac{g_1}{2}\left(\sum_{a=1}^3
  \xi_R^a\xi_L^a\right)^2+g_2\left(\sum_{a=1}^3
  \xi_R^a\xi_L^a\right)\left(\sum_{a=4}^5
  \xi_R^a\xi_L^a\right)\nonumber\\
  &+&\xi_R^6\xi_L^6\left[g_3\sum_{a=1}^3
  \xi_R^a\xi_L^a+g_4\sum_{a=4}^5 \xi_R^a
  \xi_L^a\right] +\frac{g_5}{2}\left(
  \sum_{a=4}^5 \xi_R^a\xi_L^a\right)^2
  \nonumber\\
  &+&\frac{g_6}{2}\left(\sum_{a=7}^8
  \xi_R^a\xi_L^a\right)^2 + \left(
  \xi_R^7\xi_L^7+\xi_R^8\xi_L^8\right)
  \times\nonumber\\
  &&\times\left[g_7\sum_{a=1}^3 \xi_R^a
  \xi_L^a+g_8\sum_{a=4}^5 \xi_R^a\xi_L^a
  +g_9\xi_R^6\xi_L^6\right].
  \label{Hamiltonian_tperp0_majorana}
\end{eqnarray}

The different velocities and the nine
coupling constants cannot be determined within
this phenomenological approach based on symmetries.
In this respect,  a direct standard continuum limit procedure
of the lattice model must be applied.
This is done in Appendix A and we find the
expression of the velocities:
\begin{eqnarray}
  &&v_c=v_F+\frac{a_0}{\pi}\left(
  \frac{3}{2}U-\frac{J_t}{4}
  \right)\nonumber\\
  &&v_s=v_F -\frac{a_0}{2\pi}\left(
  U-\frac{J_H}{2}+\frac{J_t}{2}\right)\nonumber\\
  &&v_t=v_F -\frac{a_0}{2\pi}\left(
  U-\frac{3J_t}{2}\right)\nonumber\\
  &&v_0=v_F -\frac{a_0}{2\pi}\left(
  U+\frac{3J_H}{2}
  +\frac{J_t}{2}\right),
\label{velo}
\end{eqnarray}
whereas the identification of the nine coupling
constants reads as follows
\begin{eqnarray}
  &&g_1=-a_0\left(U-\frac{J_H}{2}
  +\frac{J_t}{2}\right)\nonumber\\
  &&g_2=-a_0\left(U-\frac{J_H}{4}
  -\frac{J_t}{2}\right)\nonumber\\
  &&g_3=-a_0\left(U+\frac{J_H}{2}
  +\frac{J_t}{2}\right)\nonumber\\
  &&g_4=-a_0\left(U+\frac{3J_H}{4}
  -\frac{J_t}{2}\right)\nonumber\\
  &&g_5=-a_0\left(U-\frac{3J_t}{2}
  \right)\nonumber\\
  &&g_6=a_0\left(3U-\frac{J_t}{2}
  \right)\nonumber\\
  &&g_7=a_0\left(U+\frac{J_H}{4}
  -\frac{J_t}{2}\right)\nonumber\\
  &&g_8=a_0\left(U+\frac{J_t}{2}
  \right)\nonumber\\
  &&g_9=a_0\left(U-\frac{3J_H}{4}
  -\frac{J_t}{2}\right). \label{majocouplings}
\end{eqnarray}

The main advantage of this Majorana
fermions description is that
the symmetries of the original lattice model
are explicit in the low-energy effective model
(\ref{Hamiltonian_tperp0_majorana}) in sharp contrast
to the standard Abelian bosonization representation
(see for instance
Eqs.~(\ref{bosona},\ref{bosoumklapp}) of Appendix A
where the symmetries are hidden).
In particular, using Eqs.~(\ref{velo}, \ref{majocouplings}),
one can check that all extended symmetries
of Table I are indeed symmetries of
model (\ref{Hamiltonian_tperp0_majorana}).

\subsection{Duality approach}

On top of the continuous symmetries of
model (\ref{GHundmodellattice}),
the low-energy effective Hamiltonian
(\ref{Hamiltonian_tperp0_majorana}) displays
exact hidden discrete symmetries which take
the form of duality symmetries. Indeed, as
shown recently in Ref.~\onlinecite{boulat},
general weakly-interacting fermionic models
with marginal interactions exhibit
non-perturbative duality symmetries in their
low-energy description. Those will help us
to list and identify possible gapful
phases that may occur at low-energy. This
is the object of the present subsection.\\
The duality symmetries are easily identified here
 within the Majorana formalism
 (\ref{Hamiltonian_tperp0_majorana}) since they are
 built from Kramers-Wannier duality symmetries~\cite{bookboso,mussardo} of the underlying two-dimensional Ising
 models associated to the eight Majorana
 fermions $\xi^a_{R,L}$: they simply take  the
 following form $\xi^a_{L}\to -\xi^a_{L}$,
  $\xi^a_{R}\to \xi^a_{R}$. Applying the approach
  of Ref.~\onlinecite{boulat} yields 8 possible dualities,
  that can be built out of three elementary ones that one may choose as:
 \begin{eqnarray}
 \Omega_1&:&\quad \xi^{7,8}_L \to -\xi^{7,8}_L\nonumber\\
 \Omega_2&:&\quad \xi^{4,5}_L \to -\xi^{4,5}_L\nonumber\\
 \Omega_7&:&\quad \xi^{6}_L \to -\xi^{6}_L\nonumber
 \end{eqnarray}
 We now describe in detail each of the 8 phases.

\subsubsection{Spin-Peierls phase}

The essence of a duality approach is to relate
different phases between themselves.
In this respect, we need a starting phase
from which the dual phases can be obtained.
Such a phase can be most simply chosen by considering
the special fine-tuning $\Omega_0$ : $g_i=g$ ($i=1,\ldots,9$)
in Eq.~(\ref{Hamiltonian_tperp0_majorana}). The RG
study presented shortly will assess that this line is attractive under the
RG flow.
On this particular
line of the nine-dimensional parameter
space, the interacting part of the low-energy
effective model (\ref{Hamiltonian_tperp0_majorana})
takes the form of the
SO(8) Gross-Neveu model~\cite{GN}:
\begin{equation}
{\cal H}_{\rm int}^{\Omega_0} = \frac{g}{2}
\left(\sum_{a=1}^{8} \xi^a_R \xi^a_L\right)^2.
\label{SPphasemajo}
\end{equation}
The SO(8) symmetry rotates the eight Majorana
fermions and is the maximal continuous symmetry of the
interaction model
(\ref{Hamiltonian_tperp0_majorana}).
The SO(8) GN  model is integrable and a spectral gap
is generated for $g > 0$~\cite{Zamolodchikov,essler,mussardo}.
The resulting phase corresponds to a spin-Peierls
(SP) ordering with the lattice order parameter
${\cal O}_{\rm SP} =\sum_{i, l\sigma}
(-1)^i c^{\dag}_{l\sigma,i}
c^{\phantom\dag}_{l\sigma,i+1}$.

Indeed, a straightforward semiclassical
approach to model (\ref{SPphasemajo})
reveals that the bosonic fields $\Phi_{c,s,f,sf}$
of the basis (\ref{basebosons}) are pinned into
the following configurations for $g > 0$:
\begin{equation}
\langle \Phi_{a} \rangle = \sqrt{\pi} p_{a},
\qquad \langle \Phi_{a} \rangle =
\frac{\sqrt{\pi}}{2} + \sqrt{\pi} q_{a},
\label{pinningfieldSP}
\end{equation}
$p_{a}, q_{a}$ ($a=c,s,f,sf$) being integers.
In addition, the ground state degeneracy of this
 phase can be deduced within this semiclassical
  approach since there is a gauge redundancy
  in the bosonization procedure (\ref{bosofer}):
\begin{equation}
\Phi_{l\sigma R,L} \rightarrow
\Phi_{l\sigma R,L} + \sqrt{\pi} p_{l\sigma R,L},
\label{gauge}
\end{equation}
where $p_{l\sigma R,L}$ are integers ($l = 1,2$
and $\sigma=\uparrow,\downarrow$). This
transformation leaves intact the Dirac
fermion fields $R_{l\sigma}, L_{l\sigma}$.
Using the change of basis (\ref{basebosons}),
we deduce that among the field configurations
(\ref{pinningfieldSP}), only two of them are
independent (they cannot be connected by the
gauge redundancy transformation (\ref{gauge})):
\begin{eqnarray}
&&\langle \Phi_{c,s,f,sf} \rangle = 0\nonumber\\
&&\langle \Phi_{c} \rangle = \sqrt{\pi}, \qquad
\langle \Phi_{s,f,sf} \rangle = 0\,. \label{GSSP}
\end{eqnarray}
These two ground states are related by the
one-step translation symmetry $T_{a_0}$, which
is described in the bosonization approach by:
$\Phi_c \rightarrow \Phi_c + \sqrt{\pi}$. The
SP phase is thus two-fold degenerate and
spontaneously breaks the translation symmetry
$T_{a_0}$ as it should.

The continuum bosonized description of the SP
order parameter is given by:
\begin{equation}
{\cal O}_{\rm SP} \sim  \prod_{a =  c,s,f,sf}
\cos\left(\sqrt{\pi} \Phi_a\right)
+ \prod_{a= c,s,f,sf} \sin\left(\sqrt{\pi}
\Phi_a\right), \label{SPorderpara}
\end{equation}
from which we deduce that indeed this order
parameter condenses in the field configurations
 (\ref{GSSP}):
$\langle {\cal O}_{\rm SP} \rangle \ne 0$.

\subsubsection{Charge density wave phase}
Starting from the SP phase, one can infer
all possible gapful phases that may appear
at low-energy. We obtain a second degenerate
phase by performing a duality transformation
$\Omega_1:\,\xi^{7,8}_L \to -\xi^{7,8}_L$
which is a symmetry of model
(\ref{Hamiltonian_tperp0_majorana})
if $g_{7,8,9}\to -g_{7,8,9}$. The resulting
gapful phase, named ${\cal M}_1$, is governed
by the following interacting Hamiltonian,
which replaces the SO(8) line
(\ref{SPphasemajo}):
\begin{equation}
{\cal H}_{\rm int}^{\Omega_1} = \frac{g}{2}
 \left(\sum_{a=1}^{6} \xi^a_R \xi^a_L
 - \xi^7_R \xi^7_L - \xi^8_R \xi^8_L \right)^2.
  \label{CDWphasemajo}
\end{equation}
In bosonic language, the duality $\Omega_1$
only affects the charge degrees of freedom and
corresponds to a simple shift of the charge bosonic
field: $\Phi_{c L} \rightarrow \Phi_{c L}
+ \sqrt{\pi}/2$
and $\Phi_{c R} \rightarrow \Phi_{c R}$.
We deduce from Eq.~(\ref{GSSP}) that
the phase ${\cal M}_1$ is two-fold degenerate
with the two semiclassical ground states:
\begin{eqnarray}
&&\langle \Phi_{c} \rangle =
\frac{\sqrt{\pi}}{2},\;\, \qquad
\langle \Phi_{s,f,sf} \rangle = 0,
\nonumber \\
&&\langle \Phi_{c} \rangle =
\frac{3 \sqrt{\pi}}{2}, \qquad
\langle \Phi_{s,f,sf} \rangle = 0\,.
\label{GSCDW}
\end{eqnarray}
Hence, the ${\cal M}_1$ phase also breaks the
one-step translation symmetry. The
semiclassical approach enables us to identify
the ${\cal M}_1$ phase as a
CDW phase,
described by the lattice order parameter
${\cal O}_{\rm CDW} = \sum_{i, l\sigma}
(-1)^i c^{\dagger}_{l\sigma,i}
c^{\phantom\dag}_{l\sigma,i}$.
Indeed, in the bosonization description,
the CDW order parameter reads as follows:
\begin{eqnarray}
{\cal O}_{\rm CDW}&\sim&  \cos\left(\sqrt{\pi}
\Phi_c\right)\prod_{a = s,f,sf}
\sin\left(\sqrt{\pi} \Phi_a\right)\nonumber \\
&-& \sin\left(\sqrt{\pi} \Phi_c\right)
\prod_{a= s,f,sf} \cos\left(\sqrt{\pi}
\Phi_a\right),\label{CDWorderpara}
\end{eqnarray}
and it obviously condenses in the ground
state configuration (\ref{GSCDW}):
$\langle {\cal O}_{\rm CDW} \rangle \ne 0$.

\subsubsection{Orbital density wave phase}

We can define a second duality transformation
$\Omega_2:\,\xi^{4,5}_L \to -\xi^{4,5}_L$,
which is indeed a symmetry of model
(\ref{Hamiltonian_tperp0_majorana})
if $g_{2,4,8}\to -g_{2,4,8}$.
In that case,  the SO(8) line (\ref{SPphasemajo})
is replaced by:
\begin{equation}
{\cal H}_{\rm int}^{\Omega_2} = \frac{g}{2}
 \left(\sum_{a=1,2,3;6,7,8} \xi^a_R
 \xi^a_L - \xi^4_R \xi^4_L - \xi^5_R
 \xi^5_L \right)^2. \label{ODWphasemajo}
\end{equation}
The duality $\Omega_2$ affects the orbital degrees
 of freedom and is represented by a shift
  in the orbital bosonic field: $\Phi_{f L}
   \rightarrow \Phi_{f L} + \sqrt{\pi}/2$.  The
   resulting phase, ${\cal M}_2$, is a two-fold
   degenerate with ground state configurations:
\begin{eqnarray}
  &&\langle \Phi_{f} \rangle =
  \frac{\sqrt{\pi}}{2},\;\, \quad
  \langle \Phi_{c,s,sf} \rangle = 0,
  \nonumber \\
  &&\langle \Phi_{c} \rangle = \sqrt{\pi} ,\quad\langle \Phi_{f} \rangle =
  \frac{\sqrt{\pi}}{2}, \quad
  \langle \Phi_{s,sf} \rangle = 0\,.
\label{GSODW}
\end{eqnarray}
Similarly to the CDW phase, one deduces
that ${\cal M}_2$ has a long-range ODW ordering. The order
parameter is: ${\cal O}_{\rm ODW} =
\sum_{i, l\sigma} (-1)^i (-1)^{l+1}
c^{\dag}_{l\sigma,i} c^{\phantom\dag}_{l\sigma,i}$
whose bosonized form reads:
\begin{eqnarray}
{\cal O}_{\rm ODW}&\sim&
\cos\left(\sqrt{\pi} \Phi_f\right)
\prod_{a = c,s,sf} \sin\left(\sqrt{\pi}
\Phi_a\right)\nonumber \\
&-& \sin\left(\sqrt{\pi} \Phi_f\right)
\prod_{a= c,s,sf} \cos\left(\sqrt{\pi}
\Phi_a\right),\label{ODWorderpara}
\end{eqnarray}
and it obviously condenses in the ground state
configurations (\ref{GSODW}):
$\langle {\cal O}_{\rm ODW} \rangle \ne 0$.

\subsubsection{Spin Peierls-$\pi$ phase}
 The last two-fold degenerate phase,
 ${\cal M}_3$, which breaks translation
 symmetry, is obtained from the SP phase
 with help of the duality
 $\Omega_3=\Omega_1\Omega_2:\,\xi^{4,5,7,8}_L
  \to -\xi^{4,5,7,8}_L$. It is a symmetry
  of model (\ref{Hamiltonian_tperp0_majorana})
   if $g_{2,4,7,9}\to -g_{2,4,7,9}$.
   The ${\cal M}_3$ phase is governed by the
   following interacting Hamiltonian:
 \begin{equation}
{\cal H}_{\rm int}^{\Omega_3} = \frac{g}{2}
 \left(\sum_{a=1,2,3;6} \xi^a_R \xi^a_L
 - \sum_{a=4,5,7,8} \xi^a_R \xi^a_L \right)^2.
 \label{SPpiphasemajo}
\end{equation}

The order parameter characterizing the
${\cal M}_3$ phase is a Spin Peierls-$\pi$
 (SP$_\pi$) order parameter with an
 alternating dimerization profile on the
 two legs: ${\cal O}_{\rm SP_\pi} =
 \sum_{i, l\sigma} (-1)^i(-1)^{l+1}
 c^{\dagger}_{l\sigma,i}c^{\phantom\dag}_{l\sigma,i+1}$.
 Its bosonized form is:
\begin{eqnarray}
{\cal O}_{\rm SP_\pi}&\sim&  \prod_{a = c,f}
\cos\left(\sqrt{\pi} \Phi_a\right)
\prod_{a = s,sf} \sin\left(\sqrt{\pi}
\Phi_a\right)\nonumber \\
&+& \prod_{a = c,f}\sin\left(\sqrt{\pi}
 \Phi_a\right)\prod_{a= s,sf}
 \cos\left(\sqrt{\pi} \Phi_a\right).
 \label{SPpiorderpara}
\end{eqnarray}
The ground state configurations of the
SP$_\pi$ phase are given by:
\begin{eqnarray}
&&\langle \Phi_{c,f} \rangle =
\frac{\sqrt{\pi}}{2},\;\, \quad
\langle \Phi_{s,sf} \rangle = 0,
\nonumber \\
&&\langle \Phi_{c} \rangle =
\frac{3 \sqrt{\pi}}{2}, \quad
\langle\Phi_{f} \rangle =
\frac{\sqrt{\pi}}{2}, \quad
\langle \Phi_{s,sf} \rangle = 0,
\label{GSSPpi}
\end{eqnarray}
and $\langle {\cal O}_{\rm SP_\pi}
\rangle \ne 0$ in these configurations.

\subsubsection{Haldane charge phase}

So far we have considered only duality symmetries which involve an
even number of Majorana fermions.  A second class of interesting
duality symmetries, called outer dualities in Ref.~\onlinecite{boulat},
is involved when an odd number of Majorana fermions is
considered. These duality symmetries give rise to the second class of
Mott-insulating phases, the non-degenerate ones, which do not spontaneously
break the translation symmetry.  A first non-degenerate gapful
phase, named ${\cal M}_4$, is obtained from the SO(8) line
(\ref{SPphasemajo}) by the duality $\Omega_4: \xi^{6,7,8}_L \to
-\xi^{6,7,8}_L$.  Such an outer duality is a symmetry of model
(\ref{Hamiltonian_tperp0_majorana}) when $g_{3,4,7,8} \rightarrow -
g_{3,4,7,8}$. The resulting effective model for the ${\cal M}_4$ phase
is:
\begin{equation}
{\cal H}_{\rm int}^{\Omega_4} = \frac{g}{2}
\left(\sum_{a=1}^{5} \xi^a_R \xi^a_L
- \sum_{a=6}^{8} \xi^a_R \xi^a_L \right)^2.
\label{HCphasemajo}
\end{equation}

A simple semiclassical analysis of this model
shows us that the bosonic fields are
pinned in the following way:
\begin{eqnarray}
&&\langle \Phi_{c} \rangle =
\frac{\sqrt{\pi}}{2} + \sqrt{\pi} p_c,
\langle \Phi_{s,f} \rangle =\langle
\Theta_{sf} \rangle = \sqrt{\pi} p_{s,f,sf},
\nonumber \\
\label{pinningfieldHCgen}
\\
&&\langle \Phi_{c} \rangle =
\frac{3 \sqrt{\pi}}{2} + \sqrt{\pi} q_c,
\langle \Phi_{s,f} \rangle =\langle
\Theta_{sf} \rangle =
\frac{\sqrt{\pi}}{2} + \sqrt{\pi} q_{s,f,sf},
\nonumber
\end{eqnarray}
where $p_{a}, q_{a}$ ($a=c,s,f,sf$) are again integers.
Using the gauge redundancy
(\ref{gauge}), we observe that the phase
${\cal M}_4$ is indeed non-degenerate with
ground state configuration:
\begin{equation}
  \langle \Phi_{c} \rangle=\frac{\sqrt{\pi}}{2},
  \quad \langle \Phi_{s,f} \rangle = \langle
  \Theta_{sf} \rangle = 0\,. \label{GSHC}
\end{equation}

Unfortunately, the order parameter of this ${\cal M}_4$ phase cannot
be written locally in terms of the original lattice fermions. In this
low-energy procedure, it involves order and disorder operators of the
underlying two-dimensional Ising models. The situation here is similar
to the RS and RT phases of the two-leg spin-$\frac{1}{2}$
ladder~\cite{shelton}.  These gapful phases are non-degenerate and
display a hidden antiferromagnetic ordering and possibly edge states
which can be revealed through non-local string order parameters~\cite{dennijs,string2leg,kim}.  In this respect, in order to build
those operators, it is useful to introduce the following quantities in
terms of the occupation numbers $n_{l\sigma,i}$ of the original
lattice fermions:
\begin{eqnarray}
n_{c,i} &=& \frac{1}{2}
\left(n_{1\uparrow,i}+n_{1\downarrow,i}
+n_{2\uparrow,i}+n_{2\downarrow,i}\right),
\nonumber\\
n_{s,i} &=& \frac{1}{2}
\left(n_{1\uparrow,i}-n_{1\downarrow,i}
+n_{2\uparrow,i}-n_{2\downarrow,i}\right),
\nonumber\\
n_{f,i} &=& \frac{1}{2}
\left(n_{1\uparrow,i}+n_{1\downarrow,i}
-n_{2\uparrow,i}-n_{2\downarrow,i}\right),
\nonumber\\
n_{sf,i} &=& \frac{1}{2}
\left(n_{1\uparrow,i}-n_{1\downarrow,i}
-n_{2\uparrow,i}+n_{2\downarrow,i}
\right). \label{cartans}
\end{eqnarray}
We then consider two classes of string-like order parameters:
\begin{eqnarray}
{\cal O}^{\rm even}_{a,i} &=&
\cos \left( \pi \sum_{k < i}\delta
n_{a,k} \right),\nonumber \\
{\cal O}^{\rm odd}_{a,i} &=& \delta
n_{a,i} \cos \left( \pi \sum_{k < i}
\delta n_{a,k} \right),\label{strings}
\end{eqnarray}
with $\delta n_{a,i} =  n_{a,i}
- \langle n_{a,i} \rangle$ and
$a=c,s,f,sf$. The string operators
(\ref{strings}) are respectively even
or odd under the transformation
$\delta n_{a,i} \rightarrow - \delta n_{a,i}$.
The bosonization description of these
string-order parameters is cumbersome
due to the non-locality of the operators
in Eq.~(\ref{strings}).
A naive continuum expression can be derived
with help of its symmetry properties
like in the two-leg spin ladder~\cite{nakamura,kim} or the  one-dimensional
extended Bose-Hubbard model~\cite{berg}:
\begin{eqnarray}
{\cal O}^{\rm even}_{a} &\sim&
\cos \left( \sqrt{\pi} \Phi_a \right),
\nonumber \\
{\cal O}^{\rm odd}_{a} &\sim&
\sin \left( \sqrt{\pi} \Phi_a \right).
\label{stringscont}
\end{eqnarray}

We thus deduce that the odd charge string operator
displays long-range ordering in the ${\cal M}_4$ phase:
\begin{eqnarray}
&&\lim_{|i-j| \rightarrow \infty}
\langle{\cal O}^{\rm odd}_{c,i}
{\cal O}^{\rm odd}_{c,j}\rangle
\nonumber\\
&  \sim &\lim_{|x-y|\rightarrow \infty}
\langle\sin\left(\sqrt{\pi}\Phi_{c}(x)\right)
 \sin\left(\sqrt{\pi}\Phi_{c}(y)\right)
 \rangle \neq 0\,.\phantom{\quad}\label{stringLROHC}
\end{eqnarray}

Using the charge pseudo-spin operator (\ref{chargesu2}), one
immediately observes that this lattice charge string-order parameter
is equivalent to the long-range ordering (\ref{stringcharge}) obtained
within the strong-coupling approach.  We thus conclude that the ${\cal
  M}_4$ phase is a HC phase which is adia\-ba\-ti\-cally connected to the HC
of the strong-coupling approach found in the vicinity of the $J_H = -
8 U/3$ line.  This phase displays a hidden ordering, described by
Eq.~(\ref{stringLROHC}), and pseudo-spin $\frac{1}{2}$ edge states, as expected
for a Haldane phase. Those (holon) edge states carry charge but are
singlet states as far as the spin and orbital degrees of freedom are
concerned (see Appendix \ref{edgestatesapp}). This result will help us
to detect numerically the HC phase in the DMRG calculations of Section
IV.

\subsubsection{Haldane orbital phase}
The next phase, ${\cal M}_5$, is found
by applying the duality symmetries
 $\xi^{4,5,6}_L \to -\xi^{4,5,6}_L$
with $g_{2,3,8,9} \to - g_{2,3,8,9}$.
The resulting dual interacting Hamiltonian
reads
\begin{equation}
{\cal H}_{\rm int}^{\Omega_5} = \frac{g}{2}
 \left(\sum_{a=1,2,3,7,8} \xi^a_R \xi^a_L
 - \sum_{a=4}^6 \xi^a_R \xi^a_L \right)^2.
  \label{HOphasemajo}
\end{equation}
Its ground state configuration is given by:
\begin{equation}
  \langle \Phi_{f} \rangle=
  \frac{\sqrt{\pi}}{2}, \quad \langle
  \Phi_{c,s} \rangle = \langle \Theta_{sf}
   \rangle = 0\,. \label{GSHO}
\end{equation}
The physical properties of the ${\cal M}_5$ phase
are very similar to the ones for the HC phase.
The orbital degrees of freedom are central
to the ${\cal M}_5$ phase and plays the
role of the charge degrees of freedom for the HC charge.
In this respect,
the ${\cal M}_5$ phase is characterized by
the long-range order of the odd orbital
string order parameter:
\begin{eqnarray}
&&\lim_{|i-j| \rightarrow \infty}
\langle{\cal O}^{\rm odd}_{f,i}
{\cal O}^{\rm odd}_{f,j}\rangle \nonumber\\
& \sim &\lim_{|x-y|\rightarrow \infty}
\langle\sin\left(\sqrt{\pi}\Phi_{f}(x)\right)
 \sin\left(\sqrt{\pi}\Phi_{f}(y)\right)
 \rangle \neq 0\,. \phantom{\quad}\label{stringLROHO}
\end{eqnarray}

Using the orbital pseudo-spin
operator (\ref{orbitalpseudospin}),  we find that
the orbital string-order
parameter (\ref{stringLROHO}) is equivalent to the long-range ordering
(\ref{stringorbital}) obtained within the strong-coupling approach.
We thus conclude that the ${\cal M}_5$ phase
is a HO phase which is adiabatically connected to the
HO of the strong-coupling approach found in the vicinity
of the SU(2)$_o$ symmetric line $J_t = - 3 J_H/4$.
The HO phase is characterized by  pseudo-spin $\frac{1}{2}$
edge states  (see Appendix \ref{edgestatesapp}) which
carry orbital quantum number only.
The orbital edge states will be useful to reveal
numerically the HO phase by means of the DMRG
approach.

\subsubsection{Rung-triplet phase}
A new non-degenerate phase, named ${\cal M}_6$, is found
by applying the duality symmetry $\Omega_6$
$\xi^{1,2,3}_L \to - \xi^{1,2,3}_L$ to the SP phase.
The effective interacting Hamiltonian which
governs the properties of the ${\cal M}_6$ phase
is:
\begin{equation}
{\cal H}_{\rm int}^{\Omega_6} =
\frac{g}{2}  \left(
\sum_{a=4}^{8} \xi^a_R \xi^a_L
- \sum_{a=1}^{3} \xi^a_R \xi^a_L \right)^2.
\label{RTphasemajo}
\end{equation}
In the ${\cal M}_6$ phase,
the ground state configuration for the bosonic fields is:
\begin{equation}
  \langle \Phi_{s} \rangle= \langle
  \Theta_{sf} \rangle=\frac{\sqrt{\pi}}{2},
  \quad \langle \Phi_{c,f} \rangle  = 0 ,
  \label{GSRT}
\end{equation}
from which we deduce that the following
string-order parameter condenses in this phase:
\begin{eqnarray}
&&\lim_{|i-j| \rightarrow \infty}
\langle{\cal O}^{\rm odd}_{s,i}
{\cal O}^{\rm odd}_{s,j}\rangle \nonumber\\
&\sim &\lim_{|x-y|\rightarrow \infty}
\langle\sin\left(\sqrt{\pi}\Phi_{s}(x)\right)
 \sin\left(\sqrt{\pi}\Phi_{s}(y)\right)
 \rangle \neq 0\,.\phantom{\quad}\label{stringLRORT}
\end{eqnarray}
This order parameter is the standard string-order
parameter (\ref{stringRT}) of the RT phase of the
two-leg spin ladder with ferromagnetic interchain coupling.
The ${\cal M}_6$ phase is thus a RT phase with
spin-$\frac{1}{2}$ edge states~\cite{orignac}.

\subsubsection{Rung-singlet phase}
Finally, the last non-degenerate phase,
called ${\cal M}_7$, is obtained from the SO(8)
line (\ref{SPphasemajo}) by the duality symmetry
$\Omega_7:\,\xi^{6}_L \to -\xi^{6}_L$.
its effective model is:
\begin{equation}
{\cal H}_{\rm int}^{\Omega_7} =
\frac{g}{2}  \left(\sum_{a=1,a\neq 6}^8
\xi^a_R \xi^a_L - \xi^6_R \xi^6_L\right)^2.
\label{RSphasemajo}
\end{equation}
In the semiclassical approach, the ground state
configuration for the bosons is  obtained
from the SP one (\ref{GSSP}) by changing the bosonic
field $\Phi_{sf}$ into its dual $\Theta_{sf}$
in the pinning configuration:
\begin{equation}
  \langle \Phi_{c,s,f} \rangle = \langle
  \Theta_{sf} \rangle = 0 \,.\label{GSRS}
\end{equation}
In particular, one observes that no odd string-order parameters
can condense into the ${\cal M}_7$ phase.  The latter phase is the
standard RS phase of the two-leg spin ladder with
antiferromagnetic interchain coupling that we have identified in
Section II within the strong-coupling approach.  This phase has no
edge state when open-boundary conditions are considered~\cite{orignac}.

\subsection{Phase diagram}
The duality symmetry approach, discussed in the previous subsection,
predicts the emergence of eight insulating phases for two-leg
electronic ladder with the symmetry group H $=$ U(1)$_{c}$ $\times$
SU(2)$_{s}$ $\times$ U(1)$_{o}$ $\times$ Z$_2$.  However, this
approach cannot determine which phases actually appear in the phase
diagram of a particular model like (\ref{GHundmodellattice}) with
H invariance. To answer this question, we need to
perform a one-loop RG calculation of model
(\ref{Hamiltonian_tperp0_majorana}) with initial conditions
(\ref{majocouplings}) for the generalized Hund model
(\ref{GHundmodellattice}).

\subsubsection{The phases of the generalized Hund model}

\begin{figure}[h]
\centering
\includegraphics[width=0.47\textwidth]{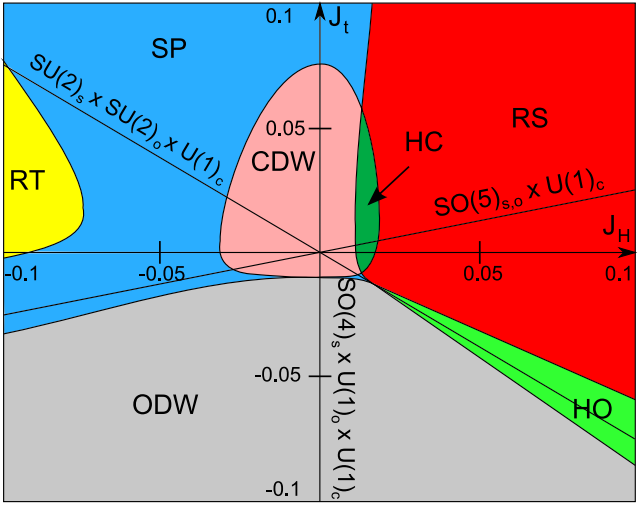}
\caption{Low-energy phase diagram of the
generalized Hund model at half-filling
for $U=-0.005t$ ($t=1$). (SP=Spin Peierls,
CDW=Charge density wave, ODW=Orbital
density wave, RS=Rung Singlet,
HC=Haldane Charge, HO=Haldane
Orbital, RT=Rung Triplet). At the intersection of the highly symmetric lines lie points with even larger symmetry (see Table I).
}\label{Hundtperp0Umin0005}
\end{figure}

The one-loop RG flow of the nine coupling constants of model
(\ref{Hamiltonian_tperp0_majorana}) can be derived by standard methods~\cite{bookboso,giamarchi}.  By neglecting the velocity anisotropy
(i.e. $v_c = v_s = v_0 = v_t = v$) and performing a suitable
redefinition of the coupling constants ($g_a = 2 \pi v f_a,
a=1,\ldots,9$), we find the one-loop RG equations:
\begin{eqnarray}
 \dot{f_1}&=&f_1^2+2f_2^2+f_3^2+2f_7^2
 \nonumber\\
 \dot{f_2}&=&2f_1f_2+f_2f_5+f_3f_4+2f_7f_8
 \nonumber\\
 \dot{f_3}&=&2f_1f_3+2f_2f_4+2f_7f_9
 \nonumber\\
 \dot{f_4}&=&f_4f_5+3f_2f_3+2f_8f_9
 \nonumber\\
 \dot{f_5}&=&3f_2^2+f_4^2+2f_8^2
 \nonumber\\
 \dot{f_6}&=&3f_7^2+2f_8^2+f_9^2
 \nonumber\\
 \dot{f_7}&=&2f_1f_7+2f_2f_8+f_3f_9+f_6f_7
 \nonumber\\
 \dot{f_8}&=&3f_2f_7+f_5f_8+f_4f_9+f_6f_8
 \nonumber\\
 \dot{f_9}&=&3f_3f_7+2f_4f_8+f_6f_9\,.
 \label{RG}
\end{eqnarray}
The nine coupling constants of the low-energy Hamiltonian
(\ref{Hamiltonian_tperp0_majorana}) are not independent: they  depend
only on the three parameters $U$, $J_h$ and $J_t$ of the original
lattice Hamiltonian (\ref{GHundmodellattice}). We thus need to use the
initial conditions (\ref{majocouplings}) to determine which
phases do come out in the zero-temperature phase diagram of the
generalized Hund model. A numerical analysis of these differential
equations, together with the results of the preceding subsection,
gives us the phase diagram of the model.  We find seven insulating
phases out of the eight possible ones found within the duality
approach.  The missing phase is the SP$_{\pi}$ phase.  Of course, by
adding next-neighbor interactions to the lattice model
(\ref{GHundmodellattice}) without breaking the symmetry group H, the
latter phase will be found.  Interestingly enough, this lattice model
with three independent interactions possess the four non-degenerate
Mott-insulating phases that we have revealed with the help the duality
symmetry and strong-coupling approaches.  In
Fig.~\ref{Hundtperp0Umin0005}, we present a section of the
three-dimensional phase diagram of the generalized Hund model at
$U=-0.005$ ($t$ is set to unity) where the seven phases appear.

\begin{figure}[h]
\centering
\includegraphics[width=0.47\textwidth]{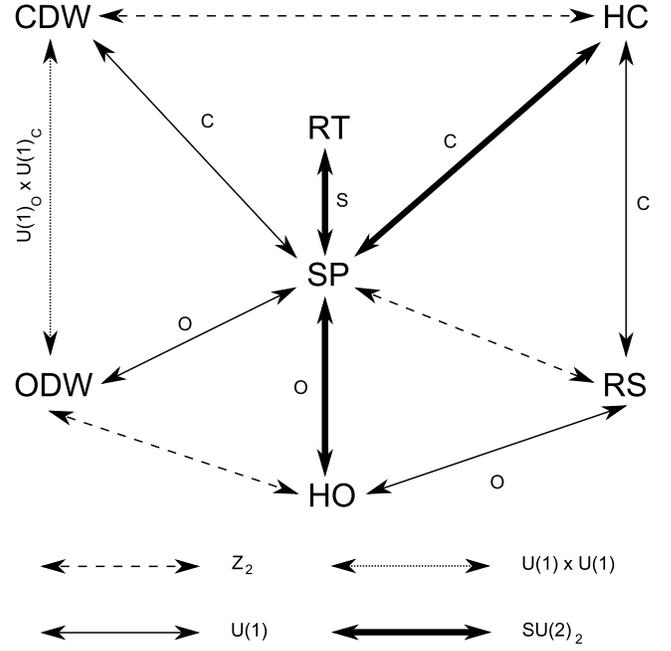}
\caption{Quantum phase transitions that can occur in the
generalized Hund model. The letter next to the arrows
indicates which degrees of freedom are critical:
c=charge, s=spin, o=orbital.}\label{phasetransitions}
\end{figure}

The duality approach that we used to obtain the
phase diagram allows for an easy characterization of
the quantum phase transitions. Those transitions are located
on the self-dual lines, where the coupling constants
that change their sign when going from one phase
to the other vanish.  Fig.~(\ref{phasetransitions}) summarizes all the phase
transitions that occur in the phase diagram of
the generalized Hund model.

We now present the zero-temperature
phase diagram of several interesting highly symmetric
models with two independent coupling constants that
we have introduced in Section II.

\subsubsection{Phase diagram of the SO(5) fermionic cold atoms model}

\begin{figure}[h]
\centering
\includegraphics[width=0.47\textwidth]{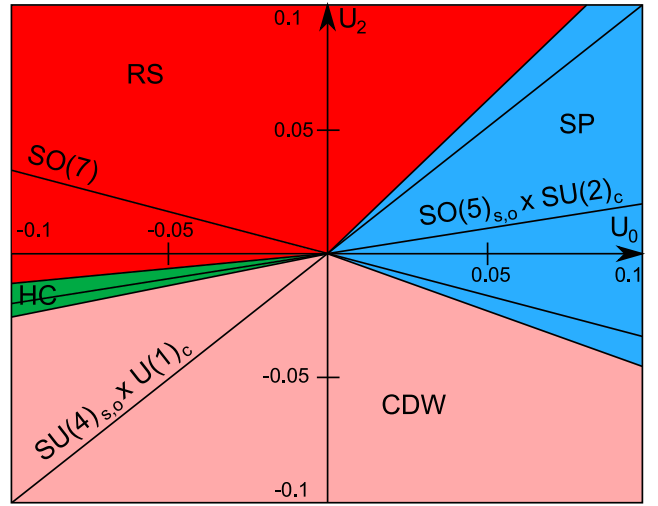}
\caption{Phase diagram of the spin-$\frac{3}{2}$ cold fermions model
at half-filling ($t=1$).}
\label{coldfermionsphasediag}
\end{figure}

Let us start with the spin-$\frac{3}{2}$ cold fermionic atoms model
(\ref{coldatoms}) which is obtained from the generalized Hund
model (\ref{GHundmodellattice}) by the fine tuning
$J_H = 4 J_t$.
In the continuum limit, the coupling constants
are naturally parametrized by the singlet and quintet
pairing $U_0$ and $U_2$:
$g_1 = g_2= g_5 =- U_0 - U_2, g_3 = g_4= U_0 -3 U_2,
g_6 = U_0 + 5 U_2, g_7 = g_8= 2 U_2,
g_9 = 2 U_0$.
The effective Majorana model (\ref{Hamiltonian_tperp0_majorana}),
describing
the physical properties of the spin-$\frac{3}{2}$ cold fermions model,
depends on five independent coupling constants.
The resulting phase diagram, as obtained from
 the one-loop RG calculation, is presented
in Fig. \ref{coldfermionsphasediag}.

\subsubsection{SO(5) SZH
model} \label{SZH}
\begin{figure}[t]
\centering
\includegraphics[width=0.47\textwidth]{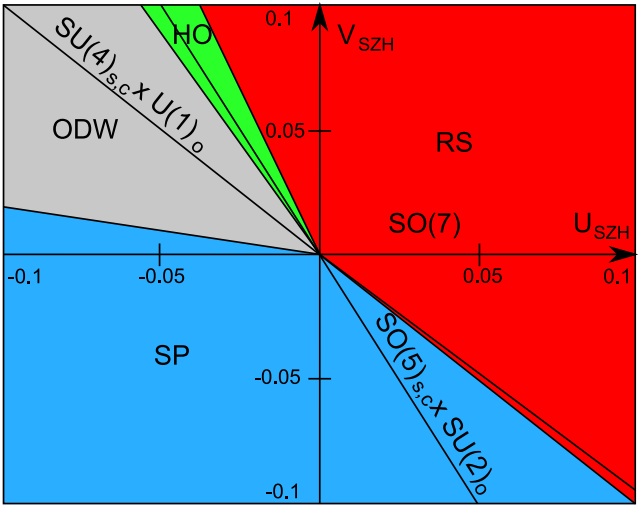}
\caption{Phase diagram of the SZH
model (\ref{SZHmodel}) with the
fine-tuning $J_{\textrm{SZH}}=
4(U_{\textrm{SZH}}+V_{\textrm{SZH}})$
at half-filling ($t=1$).}
\label{SZHphasesfinetunning2}
\end{figure}

As mentioned in the introduction,
there is a second SO(5)-symmetric model embedded in the generalized
Hund model (\ref{GHundmodellattice}):
the SZH model (\ref{SZHmodel}).
In the latter model, obtained from (\ref{GHundmodellattice})
when $J_H=8U$, the SO(5) symmetry
unifies charge and spin degrees of freedom~\cite{szh}.
The coupling constants of the
Majorana model (\ref{Hamiltonian_tperp0_majorana})
for the SZH model (\ref{SZHmodel})
read  as follows:
\begin{eqnarray}
  g_{1,6,7}&=&a_0\left(U_{\textrm{SZH}}
  +2V_{\textrm{SZH}}\right)\nonumber\\
  g_{2,8}&=&a_0 U_{\textrm{SZH}}\nonumber\\
  g_{3,9}&=&-a_0\left(3U_{\textrm{SZH}}
  +2V_{\textrm{SZH}}\right)\nonumber\\
  g_4&=&-a_0\left(3U_{\textrm{SZH}}
  +4V_{\textrm{SZH}}\right)\nonumber\\
  g_5&=&a_0\left(U_{\textrm{SZH}}
  -2V_{\textrm{SZH}}\right).
\end{eqnarray}
We obtain the
phase diagram shown in Fig.
(\ref{SZHphasesfinetunning2}).
The phases are very similar to the one
in the spin-$\frac{3}{2}$  cold fermions model
with the substitution: CDW $\to$ ODW
and HC $\to$ HO.

\subsubsection{Phase diagram of SO(4) models}
\begin{figure}[h]
\centering
\includegraphics[width=0.47\textwidth]{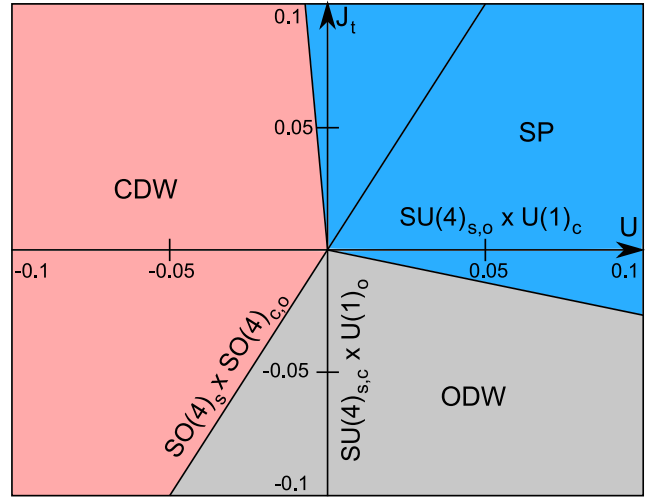}
\caption{Phase diagram of the SO(4) model  (\ref{SO4model}) with
$J_H=0$ ($t=1$) at half-filling. The continuous symmetry group is
U(1)$_c$ $\times$ U(1)$_o$ $\times$ SO(4)$_s$.}
\label{figJH0}
\end{figure}

We now turn to models which display an extended
SO(4) symmetry.
In Section II, we found two different SO(4)
models with two independent coupling constants, i.e.,
one fine-tuning with respect to the original generalized
Hund model (\ref{GHundmodellattice}).
When $J_H=0$, the lattice model (\ref{SO4model})
enjoys a U(1)$_c$ $\times$ U(1)$_o$ $\times$ SO(4)$_s$ continuous
symmetry.
The phase diagram of this model can be determined
by the low-energy approach from the identification
of the coupling constants of model (\ref{Hamiltonian_tperp0_majorana}):
\begin{eqnarray}
&& g_1=g_3=-a_0\left(U+\frac{J_t}{2}\right) \nonumber \\
&& g_2=g_4=-a_0\left(U-\frac{J_t}{2}\right)\nonumber \\
&& g_5=-a_0\left(U-\frac{3J_t}{2}\right) \nonumber \\
&& g_6=a_0\left(3U-\frac{J_t}{2}\right) \nonumber \\
&& g_7=g_9=-g_2=-g_4 \nonumber \\
&& g_8=-g_1=-g_3  .
\label{so4cont}
\end{eqnarray}
The resulting phase diagram is presented in Fig.
\ref{figJH0}. It contains an interesting line
$J_t = 2U$ with enlarged
SO(4)$_{c,o}$ $\times$ SO(4)$_s$ symmetry (see Section
\ref{sym}). It is
straightforward to see that along this line for $U<0$, where
the transition between the CDW and ODW appears, the spin degrees of freedom
are gapped while the charge and orbital degrees of
freedom are critical. Hence, the quantum phase transition
is described by a SO(4)$_1$ CFT with central charge $c=2$.

\begin{figure}[h]
\centering
\includegraphics[width=0.47\textwidth]{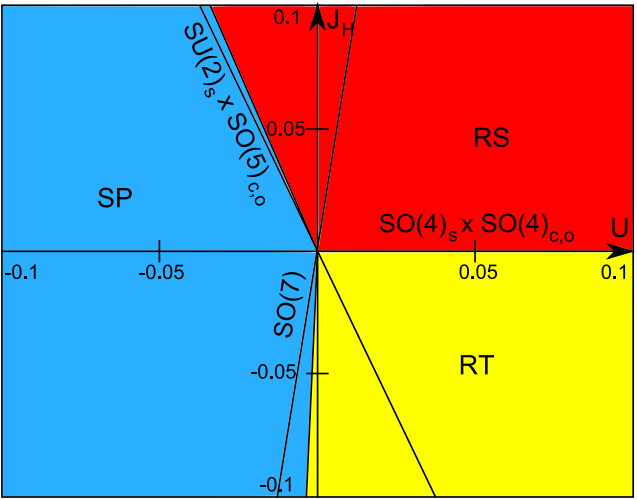}
\caption{Phase diagram of the half-filled SO(4) model with $J_t = 2U$
($t=1$)
which unifies charge and orbital degrees of freedom.}
\label{FigJt2U}
\end{figure}

The second model is defined for $J_t = 2 U$ (see Table I)
with  SO(4)$_{c,o}$ $\times$ SU(2)$_s$ continuous symmetry.
Here, the SO(4) symmetry unifies the charge and orbital
degrees of freedom.
The coupling constants of the continuum limit of
this model are given by
\begin{eqnarray}
&&g_1=-a_0\left(2U-\frac{J_H}{2}\right) \nonumber \\
&&g_2=g_7=a_0\frac{J_H}{4} \nonumber \\
&&g_3=-a_0\left(2U+\frac{J_H}{2} \nonumber \right)\\
&&g_4=g_9=-a_0\frac{3J_H}{4} \nonumber \\
&& g_5=g_6=g_8=2a_0 U ,
\label{so4biscont}
\end{eqnarray}
from which we deduce the phase diagram
of Fig. \ref{FigJt2U}.
The quantum phase transition between RS and RT phases
for $J_H =0$ and $U>0$ is now governed by the spin
degrees freedom and the SO(4)$_1$ CFT.

\subsubsection{Phase diagram of spin-orbital (charge) models}

\begin{figure}[h]
\centering
\includegraphics[width=0.47\textwidth]{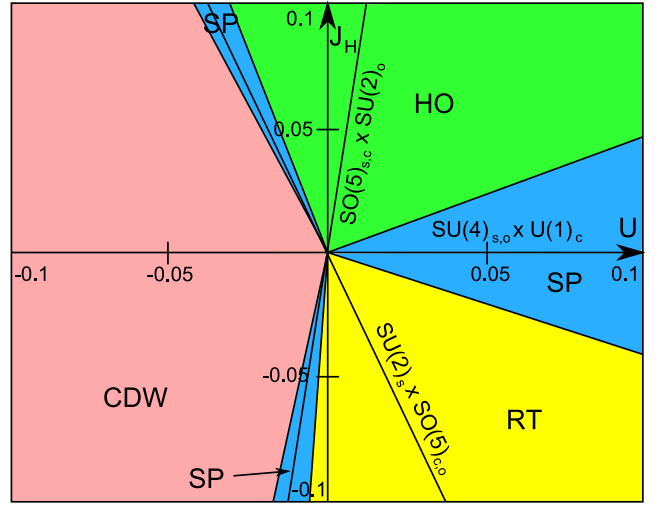}
\caption{Phase diagram of the
spin-orbital model at half-filling with
$J_t=-3J_H/4$ ($t=1$). }
\label{Jtmin3Jhsur4}
\end{figure}

The generalized Hund model (\ref{GHundmodellattice}) reduces to
the spin-orbital model when $J_t = - 3 J_H/4$ with a
U(1)$_c$ $\times$ SU(2)$_o$ $\times$ SU(2)$_s$
symmetry which has been studied in the context
of orbital degeneracy~\cite{yamashita,khomskii,mila,aza4spin,itoi00}.
The coupling constants of the continuum limit of this model
read as follows:
\begin{eqnarray}
&& g_1=-a_0\left(U-\frac{7J_H}{8}\right) \nonumber \\
&& g_2=g_3=-a_0\left(U+\frac{J_H}{8} \nonumber \right)\\
&& g_4=g_5=-a_0\left(U+\frac{9J_H}{8} \nonumber \right)\\
&& g_6=a_0\left(3U+\frac{3J_H}{8} \nonumber \right)\\
&& g_7=a_0\left(U+\frac{5J_H}{8} \nonumber \right)\\
&& g_8=g_9=a_0\left(U-\frac{3J_H}{8}\right).
\label{spinorbitalcont}
\end{eqnarray}
The phase diagram of the spin-orbital model is
presented in Fig. \ref{Jtmin3Jhsur4}.
In particular, it includes two non-degenerate Mott-insulating
phases: HO and RT phases.

\begin{figure}[h]
\centering
\includegraphics[width=0.47\textwidth]{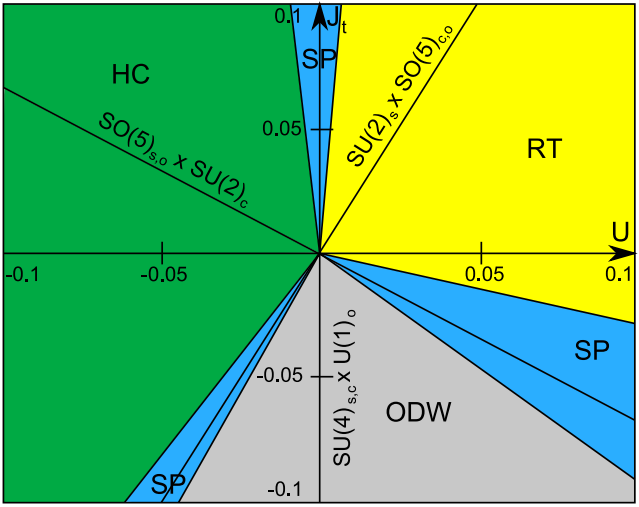}
\caption{Phase diagram of the
spin-charge model at half-filling
with $J_H=-8U/3$ ($t=1$).  The symmetry is:
  $SU(2)_s\times SU(2)_c \times U(1)_o$.}
\label{JHmin8Usur3}
\end{figure}

Finally, we consider the related spin-charge model
which is defined for the fine-tuning $J_H = - 8 U/3$
with U(1)$_o$ $\times$ SU(2)$_c$ $\times$ SU(2)$_s$ continuous
symmetry at half-filling. The resulting model is similar
to the spin-orbital model where now spin and charge are
put on the same footing.
We find the following continuum limit for the
spin-charge model:
\begin{eqnarray}
&& g_1=-a_0\left(\frac{7U}{3}+\frac{J_t}{2}\right) \nonumber \\
&& g_2=-a_0\left(\frac{5U}{3}-\frac{J_t}{2}\right) \nonumber \\
&& g_3=g_7=a_0\left(\frac{U}{3}-\frac{J_t}{2}\right) \nonumber \\
&& g_4=g_8=a_0\left(U+\frac{J_t}{2}\right) \nonumber \\
&& g_5=-a_0\left(U-\frac{3J_t}{2}\right) \nonumber \\
&& g_6=g_9=a_0\left(3U-\frac{J_t}{2}\right).
\label{spinchargecont}
\end{eqnarray}
Its phase diagram is depicted  in Fig. \ref{JHmin8Usur3}.
Here, two non-degenerate Mott insulating
phases appear: the HC and RT phases.

\subsection{Effect of
an interchain hopping term}
\label{subsectionTperp}

Let us now consider the generalized
Hund model (\ref{GHundmodellattice})
to which we add an interchain hopping
term:
\begin{equation}
  \mathcal{H}_{\bot}=-t_\bot \sum_{i,\,\sigma}
  \left(c^\dag_{1\sigma,\,i}
  c^{\phantom\dag}_{2\sigma,\,i}
   + \hc \right).
   \label{interchainhopping}
\end{equation}
This problem has been previously studied and its
phases are known~\cite{Linso8,furusaki,wu04}. We will show
that all those phases (there are eight of them)
can be connected one-by-one (by a non-trivial duality) to the phases of the generalized
Hund model (\ref{GHundmodellattice}). Our approach will use
a weak coupling analysis, and our conclusion relies on the existence
of a hidden orbital symmetry emerging at low-energy. In Appendix~\ref{appendixTperp},
we show that this
approach is also valid in the strong coupling regime, close to the
orbital line $J_t=-3J_H/4$.

In the presence of
interchain hopping, the non-interacting part of the Hamiltonian
cannot be readily diagonalized. We need
to do a change of basis and introduce
bonding and antibonding operators:
\begin{equation}
  d^{\phantom\dag}_{j\sigma}
  =\frac{1}{\sqrt{2}}\left(
  c^{\phantom\dag}_{1\sigma}+(-1)^j
  c^{\phantom\dag}_{2\sigma}\right).
  \label{bondinganttib}
\end{equation}
Using this basis, the kinetic part can be diagonalized in momentum
space; there are now two decoupled bands (bonding and antibonding
bands) and two Fermi points (provided $t_\bot<2t$) $k_{F_1}$ and
$k_{F_2}$ such that, at half-filling, $k_{F_1}+k_{F_2}=\pi/a_0$. One
could proceed by linearizing around the four Fermi points, by introducing
continuous bonding and antibonding fermionic fields $R_{a\sigma}(x)$
and $L_{a\sigma}(x)$ (with $a=1,2$, $\sigma=\uparrow,\downarrow$) by
bosonizing, refermionizing, and expanding the interactions in this
basis.  Instead, we will follow an approach that is based on the
(continuous) symmetry content of the theory, by showing that the
symmetry of the low energy, continuous theory
is essentially the \emph{same} for $t_\perp \neq 0$ as for $t_\perp = 0$.

This is not obvious: naively, the interchain hopping term breaks the
U(1)$_o$ orbital symmetry and the analysis of the preceding sections
breaks down.  However, as noticed before~\cite{Linso8,wu04}, at low
energy another U(1) symmetry emerges in the orbital sector. To see
this, let us consider the difference
$\delta=k_{F1}-k_{F2}=2\arcsin(t_\perp/2t)$. It is a continuous
function of $t_\perp$, and provided $\delta$ is not commensurate to
$\pi$, in the continuum limit and at weak coupling, one can safely
ignore umklapp terms that oscillate at wave-vector that are integer
multiples of $\delta$.  Retaining only marginal, four-fermions
interactions of the form $\prod_{i=1}^4 \Psi^{(i)}_{a_i\sigma_i}$,
with $\Psi^{(i)}=R,L$, one sees that in order for this term to be
non-oscillating and to give a contribution to the interacting part of
the continuous theory, it has to conserve \emph{separately} the
quantities $\rho_+=N_{1R}+N_{2L}$ and $\rho_-=N_{1L}+N_{2R}$, where
$N_{aL}=\sum_{\sigma}\int dx\, L^{\dagger}_{a\sigma}L^{}_{a\sigma}$
and $N_{aR}=\sum_{\sigma}\int dx\,
R^{\dagger}_{a\sigma}R^{}_{a\sigma}$ are the total number of left
(right) fermions in each of the bonding and antibonding bands.  It
results that the difference $\rho_+ -
\rho_-=(N_{1R}-N_{2R})-(N_{1L}-N_{2L})$ is conserved: this is nothing
but (twice) the total orbital current in the bonding/antibonding
basis along direction $z$. This quantity generates a U(1) orbital
symmetry $\widetilde{\mbox{U(1)}}_o$. One thus concludes that the
low-energy continuous theory has a symmetry
U(1)$_c\times$SU(2)$_s\times\widetilde{\mbox{U(1)}}_o\times Z_2$, the
same has in the $t_\perp=0$ case.

The total orbital current $\rho_+-\rho_-$ can be mapped onto the total
orbital charge (all other conserved quantities, i.e., the electric
charge and the SU(2)$_s$ spin generators, being unaffected) by the
duality $\Omega_\perp$ that leaves all Majorana fermions
$(\xi^a)_{a\neq 5}$ invariant, but changes $\xi^5_L$ to
$-\xi^5_L$. The duality $\Omega_\perp$ is highly non-local. A little
algebra shows that it has the following action on the bonding and
antibonding modes ($a=1,2$):
 \begin{equation}
  L^{\phantom\dag}_{a\,\uparrow} \to
  L^\dag_{a\,\downarrow}, \quad
  L^{\phantom\dag}_{a\,\downarrow} \to
  -L^\dag_{a\,\uparrow}.
  \label{dualitytwist}
\end{equation}

\begin{table*}[!htb]
\caption{\label{pinningstperp}Pattern of the bosonic fields pinning
for $t_\bot=0$ and $t_\bot\neq 0$.}
\begin{ruledtabular}
\begin{tabular}{lccccclccccc}
Phase &$\langle \Phi_c \rangle$ &$\langle \Phi_s \rangle$
&$\langle \Phi_f \rangle $&$\langle \Phi_{sf} \rangle $
&$\langle \Theta_{sf} \rangle $& Phase
&$\langle \Phi_c \rangle$&$\langle \Phi_s \rangle$
&$\langle \Theta_f \rangle $&$\langle \Phi_{sf} \rangle $
&$\langle \Theta_{sf} \rangle $\\
\hline
SP & $0$ & 0 & 0 & 0 & -
& S-Mott & $0$ & 0 & $\sqrt{\pi}/2$ & 0 & - \\
CDW& $\sqrt{\pi}/2$ & 0 & 0 & 0 & -
& S'-Mott & $\sqrt{\pi}/2$ & 0 & $\sqrt{\pi}/2$ & 0 & -\\
ODW& $0$ & 0 & $\sqrt{\pi}/2$ & 0 & -
& D-Mott & 0 & 0 & 0 & 0 & - \\
SP$_\pi$& $\sqrt{\pi}/2$ & 0 &
$\sqrt{\pi}/2$ & 0 & -
& D'-Mott & $\sqrt{\pi}/2$ & 0 & 0 & 0 &- \\
RS& 0 & 0 & 0 & - & 0
& CDW$_\pi$ & 0 & 0 & $\sqrt{\pi}/2$ & - & 0 \\
HC& $\sqrt{\pi}/2$ & 0 & 0 & - & 0
& PDW& $\sqrt{\pi}/2$ & 0 & $\sqrt{\pi}/2$ & - & 0 \\
HO& 0 & 0 & $\sqrt{\pi}/2$ & - & 0
& SF & 0 & 0 & 0 & - & 0\\
RT& 0 & $\sqrt{\pi}/2$ & 0 & - & $\sqrt{\pi}/2$
& FDW & 0 & $\sqrt{\pi}/2$ & $\sqrt{\pi}/2$ & -
& $\sqrt{\pi}/2$
\end{tabular}
\end{ruledtabular}
\end{table*}

With those elements at hand, the general form of the low-energy
Hamiltonian can thus be readily deduced from
(\ref{Hamiltonian_tperp0_majorana}) by performing the replacement
$\xi^5_L \to -\xi^5_L$. While this approach tells nothing about the
(bare) value of the coupling constants $g_a$, it tells us that the
structure of the RG equations (\ref{RG}) is the same, and is thus
sufficient to $(i)$ enumerate the phases of the model and $(ii)$
relate each of those phases to the phases of the generalized Hund
model (\ref{GHundmodellattice}).

From the preceding sections ($t_\bot=0$), we deduce that
there are eight possible insulating
phases. The effect of the duality
$\Omega_\bot$ is to  transform the bosonic field
$\Phi_f$ into its dual with a shift  of
$\sqrt{\pi}/2$: $\Phi_f\xrightarrow{\;\;\Omega_\bot}\Theta_f+\sqrt{\pi}/2$.
We then obtain the pinnings
of the bosonic fields from those found for
$t_\bot=0$ via the duality $\Omega_\bot$; they are summarized in
Table \ref{pinningstperp}.
 In the language of Ref.~\onlinecite{boulat},
$\Omega_\bot$ is an outer duality
(it affects only one Majorana fermion, or equivalently,
it maps a bosonic field onto its dual): it results that
it maps degenerate phases
onto non-degenerate ones and vice-versa.

The pinnings of Table \ref{pinningstperp} allow us to identify each phase with
one of the already known phases of the
generalized two-leg Hubbard ladder~\cite{Linso8,furusaki,wu04}.
The SP phase becomes an S-Mott phase with order parameter:
\begin{equation}
  {\cal O}_{\text{S-Mott}}=\sum_{i,l}c_{l\uparrow,i}
  c_{l\downarrow,i}.
\end{equation}
The CDW phase becomes the D-Mott phase, which is the same phase as RS,
and is characterized by the order parameter:
\begin{equation}
  {\cal O}_{\text{D-Mott}}=\sum_{i}c_{1\uparrow,i}
  c_{2\downarrow,i}-c_{2\uparrow,i}c_{1\downarrow,i}.
\end{equation}
The ODW (respectively SP$_\pi$) phase give an S'-Mott
(respectively D'-Mott) state
which differs from the S-Mott (respectively D-Mott) only in the pinning
of the charge bosonic field $\Phi_c$
(see Table \ref{pinningstperp})~\cite{furusaki}. Such order parameters have
the slowest decaying correlation function
when the system is doped.
The non-degenerate RS phase will become the two-fold degenerate
CDW$_{\pi}$ phase with an inter-leg phase
difference~\cite{comment}:
\begin{equation}
  {\cal O}_{\text{CDW-}\pi} = \sum_{i, l\sigma}
(-1)^i (-1)^{l+1} c^{\dagger}_{l\sigma,i}
c^{\phantom\dag}_{l\sigma,i} .
\end{equation}
The HC phase becomes a p-density wave (PDW) phase
which is described by the condensation of
the order parameter~\cite{comment2}:
\begin{equation}
{\cal O}_{\text{PDW}} =\sum_{i, l\sigma}
(-1)^i(-1)^{l+1} \left(c^{\dag}_{l\sigma,i}
c^{\phantom\dag}_{l\sigma,i+1}+\hc\right) .
\end{equation}
The HO phase gives a staggered flux (SF) phase (or a
d-density wave phase) whose ground states display
currents circulating around a plaquette, with order parameter:
\begin{eqnarray}
{\cal O}_{\text{SF}} &=&i\sum_{i,\sigma}
(c^\dag_{1\sigma,i}c^{\phantom \dag}_{2\sigma,i}
+c^{\dag}_{2\sigma,i}c^{\phantom \dag}_{2\sigma,i+1}\nonumber\\
&+&c^{ \dag}_{2\sigma,i+1}c^{\phantom \dag}_{1\sigma,i+1}
+c^{\dag}_{1\sigma,i+1}c^{\phantom \dag}_{1\sigma,i}
-\hc) .
\end{eqnarray}
This phase spontaneously breaks  time-reversal symmetry.
Finally, the RT phase will become a f-density
wave (FDW). This phase consists in currents flowing
along the diagonals of plaquettes:
\begin{eqnarray}
{\cal O}_{\text{FDW}} &=&i\sum_{i,\sigma}
(c^\dag_{2\sigma,i+1}c^{\phantom \dag}_{1\sigma,i}
-c^{\dag}_{1\sigma,i}c^{\phantom \dag}_{2\sigma,i+1}\nonumber\\
&+&c^{ \dag}_{1\sigma,i+1}c^{\phantom \dag}_{2\sigma,i}
-c^{\dag}_{2\sigma,i}c^{\phantom \dag}_{1\sigma,i+1}).
\end{eqnarray}
It also breaks time-reversal symmetry.

\begin{figure}[h]
\centering
\includegraphics[width=0.47\textwidth]{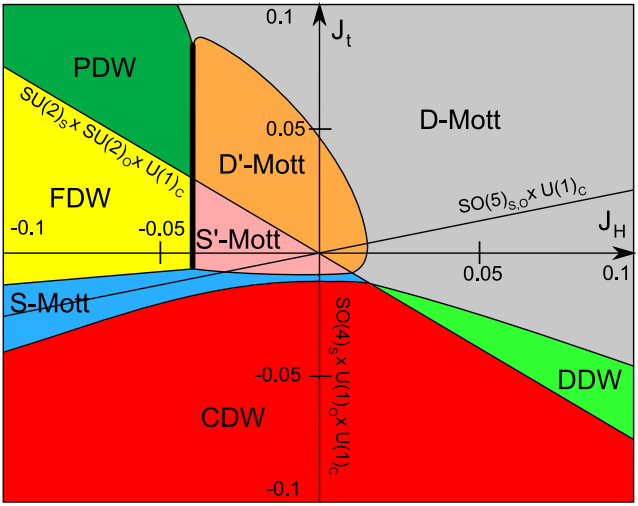}
\caption{The eight fully gapped phases of the generalized
Hund chain with a transverse hopping at half-filling ($U = - 0.005t$).}
\label{phasetperp}
\end{figure}

As already emphasized, the emergent duality symmetry $\Omega_\bot$
provides only a correspondence between the set of phases
with  $t_{\perp} = 0$
with those of $t_{\perp} \ne 0$. By no means, it maps
a given model defined by $(t_{\perp} = 0, U, J_H, J_t)$
onto the model $(t_{\perp} \ne 0, U, J_H, J_t)$.
In this respect, it is necessary to use the one-loop RG calculation
to map out the phase diagram of the generalized Hund chain
with a transverse hopping. Using Eq.~(\ref{RG}) and
the initial conditions  for that model
\begin{eqnarray}
  &&g_1=a_0\left(-U+\frac{J_H}{2}
  -\frac{J_t}{2}\right)\nonumber\\
  &&g_2=a_0\left(-\frac{3J_H}{8}
  -\frac{J_t}{2}\right)\nonumber\\
  &&g_3=a_0\left(-U+\frac{J_H}{4}
  +\frac{J_t}{2}\right)\nonumber\\
  &&g_4=g_2\nonumber\\
  &&g_5=a_0\left(U+\frac{3J_H}{4}
  -\frac{J_t}{2}\right)\nonumber\\
  &&g_6=a_0\left(3U
  -\frac{J_t}{2}\right)\nonumber\\
  &&g_7=a_0\left(U+\frac{J_H}{4}
  -\frac{J_t}{2}\right)\nonumber\\
  &&g_8=g_2\nonumber\\
  &&g_9=a_0\left(U+\frac{J_t}{2}
  \right),
\label{couplingstperp}
\end{eqnarray}
we find the presence of the eight insulating phases.
Figure \ref{phasetperp} presents a section
of the phase diagram for $U = -0.005t$ where
the eight phases are revealed.
Interestingly enough, the generalized Hund chain model
with a transverse hopping turns out to be the minimal model for two-leg
electronic ladder models
which contains the eight insulating phases found
within the low-energy approach~\cite{furusaki,wu04}.
In particular, one does not need to add further interactions
(first neighbors density interactions for instance)
contrarily to Refs.~\onlinecite{furusaki,wu04}.

\section{DMRG calculations}

We now carry out numerical calculations using DMRG in order to
investigate the various phase diagrams. For simplicity, we will
restrict ourselves to the $t_\perp=0$ case for which we can use a
purely one-dimensional implementation of the model. Moreover, for a system
of finite size $L$, we can
fix the total number of particles $N=2Q_c=2\sum_i n_{c,i}=2L$, the $z$-component of the total spin
$S^z= \sum_i n_{s,i}$ as well as
the $z$-component of the total orbital operator $T^z = \sum_i n_{f,i}$
(see Eq.~(\ref{cartans})), so that the states are labeled by the triplet $(Q_c,S^z,T^z)$. Typically,
we keep up to 1600 states, which allow to
have an error below $10^{-6}$, and we use open boundary
conditions (OBC).

\begin{figure}[!h]
\centering
\includegraphics[width=0.47\textwidth,clip]{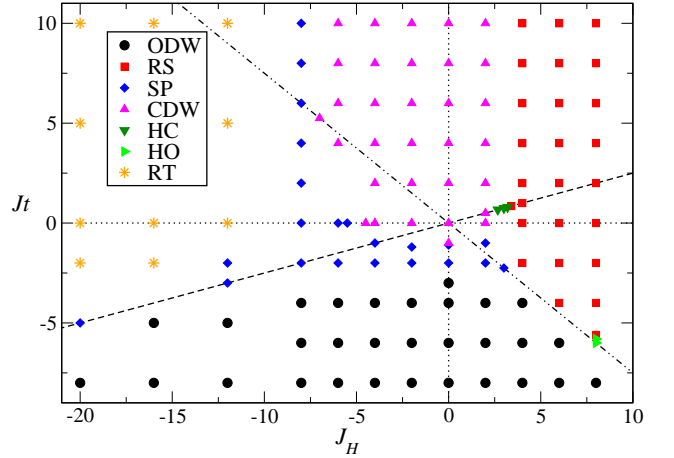}
\caption{Numerical phase diagram of the
generalized Hund model at half-filling
for $U=-t$. Notations are the same as in Fig.~\ref{Hundtperp0Umin0005}.
}\label{fig1_dmrg}
\end{figure}

As it has been revealed by the low-energy approach, there are eight
possible insulating phases, that are related by duality
transformations. In order to characterize them, we can either compute
local quantities (bond kinetic energy, local density, \ldots) to
identify states that break a given symmetry, or investigate the
presence or absence of various edge states to detect non-degenerate Mott
phases.  Thus, the three possible Haldane phases that have been
predicted (RT, HC, and HO) can be characterized by looking for the
presence of edge states with quantum numbers $(Q_c,S_z,T_z)$
respectively equal to $(L,1,0)$, $(L+1,0,0)$ and $(L,0,1)$ (see
details in Appendix~\ref{edgestatesapp}).

\begin{figure}[h]
\centering
\includegraphics[width=0.48\textwidth,clip]{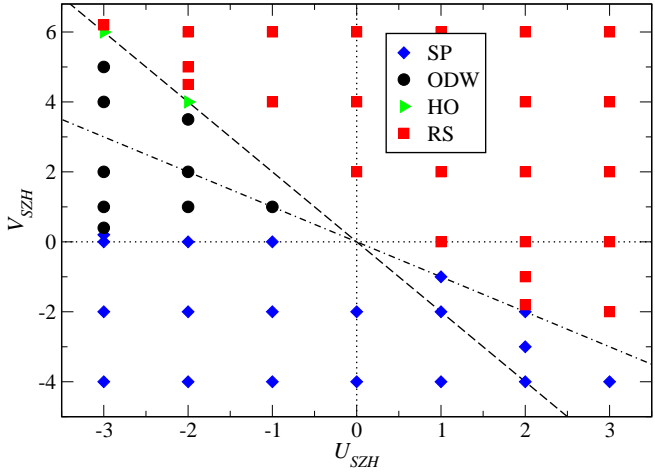}
\caption{Numerical phase diagram of the SZH SO(5) model with the fine-tuning
  $J_{SZH}=4(U_{SZH}+V_{SZH})$ at half-filling. Notations are the same
  as in Fig.~\ref{SZHphasesfinetunning2}. Dashed lines indicate models
  with higher symmetry (see Table 1).  }\label{fig2_dmrg}
\end{figure}

From these measurements, we can draw various cuts of the phase diagram
in the $(U,J_H,J_t)$ parameter space. In Fig.~\ref{fig1_dmrg}, we
present our data for fixed $U=-t$. Seven (out of eight) insulating
phases are found and the overall topology nicely agrees with the
low-energy predictions shown in Fig.~\ref{Hundtperp0Umin0005}, although
those were obtained with a much weaker interaction
$U=-0.005t$.
As another example, we consider a highly symmetric case, namely the
 SZH
model (see Section ~\ref{SZH}),
which corresponds to the case $J_H=8U$.
We recall that the low-energy
phase diagram of this model is shown in Fig.~\ref{SZHphasesfinetunning2}.
The numerical phase diagram (Fig.~\ref{fig2_dmrg}) is in
excellent agreement not only for
the overall topology but also for the location of
the phase boundaries.

\begin{figure}[!h]
\centering
\includegraphics[width=0.48\textwidth,clip]{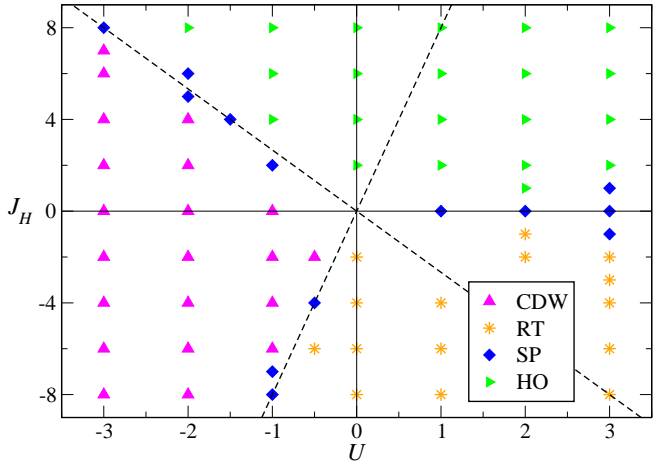}
\caption{Numerical phase diagram of the
spin-orbital model ($J_t=-3J_H/4$) at half-filling.
Notations are the same as in Fig.~\ref{Jtmin3Jhsur4}. Dashed lines indicate
models with higher symmetry (see Table 1).
}\label{fig3_dmrg}
\end{figure}

Finally, we also did simulations
for the spin-orbital model obtained by fixing $J_t=-3J_H/4$.
In that case as well, we observe that the low-energy
predictions (see Fig.~\ref{Jtmin3Jhsur4})
and our numerical data (see Fig. \ref{fig3_dmrg}) coincide very well.

In brief, for all the cases we considered when
$t_\perp=0$, we obtain an excellent agreement between the low-energy
predictions (obtained for very weak couplings) and the
numerical phase diagram (obtained at moderate couplings). This is an important
result of our paper.

The case of a transverse hopping $t_{\perp} \ne 0$ is much
more difficult to analyze with respect to the low-energy approach.
As discussed in Section III D, in the latter approach,
there is an emergent
$\widetilde{\mbox{U(1)}}$ symmetry in the orbital space which is not
present on the lattice. It means that we need to consider long-size systems
in the DMRG calculations in order to
compare with the low-energy predictions.
A second difficulty arises with respect to the numerical discrimination
between the S-Mott (respectively D-Mott) phase and
the S'-Mott (respectively D'-Mott) phase.
A numerical analysis of relevant string-order parameters
is clearly called for to distinguish them.
To the best of our knowledge, we are not aware of
any numerical study which reports the
existence of the (S,D)'-Mott phases in generalized two-leg electronic
ladders.
We plan to investigate elsewhere
the numerical phase diagram of the generalized Hund
model with a transverse hopping.

\section{Concluding Remarks}

In summary, we have investigated
the phase diagram of the generalized Hund model (\ref{GHundmodellattice})
with global symmetry group
H $=$ U(1)$_{c}$ $\times$ SU(2)$_{s}$ $\times$
U(1)$_{o}$ $\times$ Z$_2$
at half-filling
by means of complementary low-energy and DMRG techniques.
The interest of this model is manifold. First, it covers and unifies
several highly symmetric lattice models: the SO(5)-symmetric model
describing spin-$\frac{3}{2}$ cold fermions; another (different) SO(5)-symmetric model
introduced by Scalapino, Zhang and Hanke that unifies antiferromagnetism
and d-wave superconductivity; and two SU(2)$\times$SU(2)-symmetric
spin-orbital models.
While its orbital U(1) symmetry can seem a little odd from the point
of view of applications to the description of Hubbard two-leg ladders, we show
that at weak-coupling, it shares the same continuous low-energy effective theory
with the well known Hubbard two-leg ladder with interchain hopping.
Third, it is directly relevant to the description of Ytterbium 171 loaded
into an optical 1D trap.
We are able to treat on an equal footing all the phases
appearing in those models coming from different contexts.
The  rich phase diagram of the generalized Hund model has to be contrasted with the minimal character
and simplicity of this model, which only depends on three microscopic couplings.

We briefly recall our main results: by means of a duality approach, we predict that the phase diagram
for half-filled four-component
fermions with global symmetry H $=$ U(1)$_{c}$ $\times$ SU(2)$_{s}$ $\times$
U(1)$_{o}$ $\times$ Z$_2$ consists in eight Mott-insulating phases.
These phases fall into two different
classes. A first class consists of two-fold degenerate
fully gapped density phases which spontaneously break
a discrete symmetry present on the lattice.
The second class comprises four non-degenerate Mott-insulating phases
which are characterized by
non-local string order parameters: RS, RT, HC, and HO phases.
A one-loop RG calculation for model (\ref{GHundmodellattice})
reveals the existence of seven phases out of the eight ones
consistent with the duality approach.
The missing phase, which can be found by adding
further nearest-neighbor interactions, is
an alternating bond ordered phase (SP$_{\pi}$).
These results have been confirmed numerically by means
of a DMRG approach for moderate couplings.
In this respect, we found an excellent agreement
between the two complementary approaches.

Finally, we have connected our low-energy results
to the eight previously known insulating phases
found in generalized two-leg ladders with a transverse
hopping $t_{\perp}$ term.
When $t_{\perp} \ne 0$, the U(1) orbital symmetry
is lost on the lattice but becomes an emergent symmetry
at low-energy.
The duality approach with global symmetry group H
can then still be applied in the presence of a
interchain hopping $t_{\perp}$ term.
In this respect, we discovered a non-local duality
which maps the eight Mott-insulating phases for $t_{\perp} =0$
onto the eight phases previously known for $t_{\perp} \ne 0$.
The one-loop RG approach to the generalized Hund model
with interchain hopping predicts the stabilization of
the eight Mott-insulating phases.
The latter model is thus the minimal model for two-leg electronic ladders
which displays the eight Mott-insulating phases at weak coupling.
We also discuss the fate of this emergent U(1) orbital symmetry
when going from the weak-coupling to the strong coupling regime.
We show that the weak- and strong-coupling phases coincide in the vicinity of
the orbital symmetric line, but our analysis suggests a breakdown of this
picture away from this special line.

We hope that future experiments on
$^{171}$Yb or alkaline-earth cold fermions atoms
loaded into a 1D optical lattice will reveal some
of the exotic insulating phases found in our study.

\acknowledgments
We would like to thank P. Azaria and G. Roux for collaborations
on related problems.
P.L. is very grateful to the department of physics, university of Gothenburg
for hospitality during the completion of this work.
S.C. thanks CALMIP for allocation of CPU time.

\appendix

\section{Continuum Limit}
In this Appendix, we present the technical details
of the continuum limit of the generalized Hund
model (\ref{GHundmodellattice}).
The starting point of the low-energy approach is
the linearization around the Fermi points $\pm k_F$
of the dispersion relation for non-interacting
fermions. Four
left and right moving Dirac fermions
$L_{l\sigma}, R_{l\sigma}$ ($l=1,2$ and
$\sigma=\uparrow,\downarrow$) are then introduced to describe
the lattice fermions $c_{l\sigma,i}$ in the
continuum limit (see Eq.~(\ref{contfer})).

The next step of the approach is the introduction
of four chiral bosonic fields $\Phi_{l\sigma R,L}$
 through
the Abelian bosonization of Dirac fermions~\cite{bookboso,giamarchi}:
\begin{eqnarray}
R_{\,l\sigma}&=&\frac{\kappa_{l\sigma}}
{\sqrt{2\pi a_0}}\exp{\left(i\sqrt{4\pi}
\Phi_{l\sigma R}\right)}\nonumber\\
  L_{l\sigma}&=&\frac{\kappa_{l\sigma}}
  {\sqrt{2\pi a_0}}\exp{\left(-i\sqrt{4\pi}
  \Phi_{l\sigma L}\right)},
\label{bosofer}
\end{eqnarray}
where the bosonic fields satisfy the following
commutation
relation:
\begin{equation}
\left[\Phi_{l\sigma R}, \Phi_{l'\sigma' L}\right]
= \frac{i}{4}\delta_{l l'}\delta_{\sigma \sigma'}.
\label{commutator}
\end{equation}
The presence of the Klein factors
$\kappa_{l\sigma}$ ensures the correct
anticommutation of the fermionic operators.
The Klein factors satisfy the anticommutation
rule $\{\kappa_{l\sigma},\kappa_{l'\sigma'}\}
=2\delta_{l l'}\delta_{\sigma \sigma'}$ and they
are constrained so that $\Gamma^2 = 1$, with
$\Gamma=\kappa_{1\uparrow}\kappa_{1\downarrow}
\kappa_{2\uparrow}\kappa_{2\downarrow}$.
Hereafter, we will work within the $\Gamma =1$
sector.
It will be convenient to work with a pair of
dual non-chiral bosonic fields:
$\Phi_{l\sigma}=\Phi_{ l\sigma L}
+\Phi_{l\sigma R}$ and $\Theta_{l\sigma}
=\Phi_{l\sigma L}-\Phi_{l\sigma R}$.
 Last, let us introduce a SU(4) basis that will
 allow us to separate charge and non-Abelian
 (spin) degrees of freedom:
\begin{eqnarray}
  &&\Phi_{1\uparrow}=\frac{1}{2}
  (\Phi_c+\Phi_s+\Phi_f+\Phi_{sf})\nonumber\\
  &&\Phi_{1\downarrow}=\frac{1}{2}
  (\Phi_c-\Phi_s+\Phi_f-\Phi_{sf})\nonumber\\
  &&\Phi_{2\uparrow}=\frac{1}{2}
  (\Phi_c+\Phi_s-\Phi_f-\Phi_{sf})\nonumber\\
  &&\Phi_{2\downarrow}=\frac{1}{2}
  (\Phi_c-\Phi_s-\Phi_f+\Phi_{sf}).
  \label{basebosons}
\end{eqnarray}

In sharp contrast with incommensurate fillings,
there is no spin charge separation at half-filling.
 Indeed, in this special case, chiral umklapp
 processes couple those degrees of freedom.
 Consequently, the resulting low-energy
 Hamiltonian corresponding to model
 (\ref{GHundmodellattice}) takes the form:
 $\mathcal{H}=\mathcal{H}_c+\mathcal{H}_{s}
 +\mathcal{H}_{\textrm{umklapp}}$.

The charge degrees of freedom are described by:
\begin{equation}
 \mathcal{H}_c=\frac{v_F}{2}
  \left[(\partial_x \Phi_c)^2+(\partial_x \Theta_c)^2
  \right]+ \Delta v_c\, (\partial_x \Phi_c)^2,
  \label{bosocharge}
\end{equation}
where $v_F=a_0 t$ is the Fermi velocity.

The part of the bosonized Hamiltonian
corresponding to the non-abelian degrees
of freedom is:
\begin{eqnarray}
  \mathcal{H}_s&=&\frac{v_F}{2} \sum_{a=s,f,sf}
  \left[(\partial_x \Phi_a)^2
  +(\partial_x \Theta_a)^2\right] \nonumber\\
   &+&\sum_{a=s,f,sf} \Delta v_a \,
   (\partial_x \Phi_a)^2\nonumber\\
  &+&A_1 \left(C_{sfR}^{\sqrt{16\pi}}
  +C_{sfL}^{\sqrt{16\pi}}\right)\nonumber\\
  &+&A_2 \left(C_{s}^{\sqrt{4\pi}}
  +C_{f}^{\sqrt{4\pi}}\right)
  \tilde{C}_{sf}^{\sqrt{4\pi}}
  + A_3 \,C_{s}^{\sqrt{4\pi}}
  C_{f}^{\sqrt{4\pi}} \nonumber\\
  &+& A_4 \,C_{s}^{\sqrt{4\pi}}
  C_{sf}^{\sqrt{4\pi}}
  + A_5 \,C_{f}^{\sqrt{4\pi}}
  C_{sf}^{\sqrt{4\pi}}, \label{bosona}
\end{eqnarray}
where we used the compact notation:
$C_a^\beta=\cos{(\beta \Phi_a)}$, and
$\tilde{C}_a^\beta=\cos{(\beta \Theta_a)}$.
The bare parameters are:
\begin{eqnarray}
  &&\Delta v_c=\frac{a_0}{\pi}
  \left(\frac{3 U}{2}
  -\frac{J_t}{4}\right)\nonumber\\
  &&\Delta v_s=-\frac{a_0}{\pi}
  \left(\frac{ U}{2}-\frac{J_H}{4}
  +\frac{J_t}{4}\right)\nonumber\\
  &&\Delta v_f=-\frac{a_0}{\pi}
  \left(\frac{ U}{2}-\frac{3J_t}{4}
  \right)\nonumber\\
  &&\Delta v_{sf}=-\frac{a_0}{\pi}
  \left(\frac{U}{2}+\frac{J_H}{4}
  +\frac{J_t}{4}\right)\nonumber\\
  &&A_1=-\frac{J_H a_0}{4\pi^2}\nonumber\\
  &&A_2=-\frac{J_H a_0}{2\pi^2}\nonumber\\
  &&A_3=\frac{a_0}{\pi^2}\left(U
  -\frac{J_H}{4}-\frac{J_t}{2}\right)\nonumber\\
  &&A_4=\frac{a_0}{\pi^2}\left(U
  +\frac{J_t}{2}\right)\nonumber\\
  &&A_5=\frac{a_0}{\pi^2}\left(U
  +\frac{J_H}{4}-\frac{J_t}{2}\right).
\end{eqnarray}
Finally, the umklapp part of the Hamiltonian
reads:
\begin{eqnarray}
  \mathcal{H}_{\textrm{umklapp}}&=&A_6
  \,C_{c}^{\sqrt{4\pi}}C_{s}^{\sqrt{4\pi}}
  + A_7 \,C_{c}^{\sqrt{4\pi}} C_{f}^{\sqrt{4\pi}}
  \nonumber\\
  &+& A_8 \,C_{c}^{\sqrt{4\pi}}
   C_{sf}^{\sqrt{4\pi}} + A_9 \,C_{c}^{\sqrt{4\pi}}
   \tilde{C}_{sf}^{\sqrt{4\pi}}, \label{bosoumklapp}
\end{eqnarray}
with
\begin{eqnarray}
  &&A_6=\frac{a_0}{\pi^2}\left(-U
  -\frac{J_H}{4}+\frac{J_t}{2}\right)\nonumber\\
  &&A_7=\frac{a_0}{\pi^2}\left(-U
  -\frac{J_t}{2}\right)\nonumber\\
  &&A_8=\frac{a_0}{\pi^2}\left(-U
  +\frac{J_H}{4}+\frac{J_t}{2}\right)\nonumber\\
  &&A_9=A_2=-\frac{J_H a_0}{2\pi^2}\,.
\end{eqnarray}

In the end, a refermionization procedure
will allow us to make the exact
U(1)$_c$ $\times$ SU(2)$_s$ $\times$ U(1)$_o$  $\times$ Z$_2$
continuous symmetry explicit in the
effective Hamiltonian. For this purpose, we
introduce eight left and right moving Majorana
fermions through:
\begin{eqnarray}
  &&\xi_L^2+i\xi_L^1=
  \frac{\eta_1}{\sqrt{\pi a_0}}
  \exp{(-i\sqrt{4\pi}\Phi_{sL})}\nonumber\\
  &&\xi_R^2+i\xi_R^1=
  \frac{\eta_1}{\sqrt{\pi a_0}}
  \exp{(i\sqrt{4\pi}\Phi_{sR})}\nonumber\\
  &&\xi_L^4-i\xi_L^5=
  \frac{\eta_2}{\sqrt{\pi a_0}}
  \exp{(-i\sqrt{4\pi}\Phi_{fL})}\nonumber\\
  &&\xi_R^4-i\xi_R^5=
  \frac{\eta_2}{\sqrt{\pi a_0}}
  \exp{(i\sqrt{4\pi}\Phi_{fR})}\nonumber\\
  &&\xi_L^6+i\xi_L^3=
  \frac{\eta_3}{\sqrt{\pi a_0}}
  \exp{(-i\sqrt{4\pi}\Phi_{sfL})}\nonumber\\
  &&\xi_R^6+i\xi_R^3=
  \frac{\eta_3}{\sqrt{\pi a_0}}
  \exp{(i\sqrt{4\pi}\Phi_{sfR})}\nonumber\\
  &&\xi_L^8+i\xi_L^7=
  \frac{\eta_4}{\sqrt{\pi a_0}}
  \exp{(-i\sqrt{4\pi}\Phi_{cL})}\nonumber\\
  &&\xi_R^8+i\xi_R^7=
  \frac{\eta_4}{\sqrt{\pi a_0}}
  \exp{(i\sqrt{4\pi}\Phi_{cR})},\label{refer}
\end{eqnarray}
where $\eta_{1,2,3,4}$ are again Klein factors
 which ensure the adequate anticommutation rules
 for the fermions. Using this correspondence
 rules, equations (\ref{bosocharge},
 \ref{bosona},\ref{bosoumklapp}) can be
 expressed in terms of these eight Majorana
 fermions. We thus finally obtain the
 low-energy effective theory for the
 generalized Hund model (\ref{GHundmodellattice})
 at half-filling:
\begin{eqnarray}
  \mathcal{H}&=& - \frac{i v_c}{2}\sum_{a=7}^8
  (\xi_R^a \partial_x \xi_R^a
  - \xi_L^a \partial_x \xi_L^a)\nonumber\\
  &-&\frac{i v_s}{2}\sum_{a=1}^3 (\xi_R^a
  \partial_x \xi_R^a - \xi_L^a \partial_x
  \xi_L^a)\nonumber\\
  &-&\frac{i v_t}{2}\sum_{a=4}^5 (\xi_R^a
  \partial_x \xi_R^a - \xi_L^a \partial_x
  \xi_L^a)\nonumber\\
  &-&\frac{i v_0}{2} (\xi_R^6 \partial_x
  \xi_R^6 - \xi_L^6 \partial_x \xi_L^6)\nonumber\\
  &+&\frac{g_1}{2}\left(\sum_{a=1}^3
  \xi_R^a\xi_L^a\right)^2+g_2\left(\sum_{a=1}^3
  \xi_R^a\xi_L^a\right)\left(\sum_{a=4}^5
  \xi_R^a\xi_L^a\right)\nonumber\\
  &+&\xi_R^6\xi_L^6\left[g_3\sum_{a=1}^3
  \xi_R^a\xi_L^a+g_4\sum_{a=4}^5 \xi_R^a
  \xi_L^a\right] +\frac{g_5}{2}\left(
  \sum_{a=4}^5 \xi_R^a\xi_L^a\right)^2
  \nonumber\\
  &+&\frac{g_6}{2}\left(\sum_{a=7}^8
  \xi_R^a\xi_L^a\right)^2 + \left(
  \xi_R^7\xi_L^7+\xi_R^8\xi_L^8\right)
  \times\nonumber\\
  &&\times\left[g_7\sum_{a=1}^3 \xi_R^a
  \xi_L^a+g_8\sum_{a=4}^5 \xi_R^a\xi_L^a
  +g_9\xi_R^6\xi_L^6\right],
  \label{Hamiltonian_tperp0_majoranapp}
\end{eqnarray}
where the different velocities and
couplings are given by:
\begin{eqnarray}
  &&v_c=v_F+\frac{a_0}{\pi}\left(
  \frac{3}{2} U-\frac{J_t}{4}
  \right)\nonumber\\
  &&v_s=v_F -\frac{a_0}{2\pi}\left(
  U-\frac{J_H}{2}+\frac{J_t}{2}\right)\nonumber\\
  &&v_t=v_F -\frac{a_0}{2\pi}\left(
  U-\frac{3J_t}{2}\right)\nonumber\\
  &&v_0=v_F -\frac{a_0}{2\pi}\left(
  U+\frac{3J_H}{2}
  +\frac{J_t}{2}\right)\nonumber\\
  &&g_1=-a_0\left(U-\frac{J_H}{2}
  +\frac{J_t}{2}\right)\nonumber\\
  &&g_2=-a_0\left(U -\frac{J_H}{4}
  -\frac{J_t}{2}\right)\nonumber\\
  &&g_3=-a_0\left(U +\frac{J_H}{2}
  +\frac{J_t}{2}\right)\nonumber\\
  &&g_4=-a_0\left(U +\frac{3J_H}{4}
  -\frac{J_t}{2}\right)\nonumber\\
  &&g_5=-a_0\left(U -\frac{3J_t}{2}
  \right)\nonumber\\
  &&g_6=a_0\left(3 U-\frac{J_t}{2}
  \right)\nonumber\\
  &&g_7=a_0\left(U +\frac{J_H}{4}
  -\frac{J_t}{2}\right)\nonumber\\
  &&g_8=a_0\left(U +\frac{J_t}{2}
  \right)\nonumber\\
  &&g_9=a_0\left(U  - \frac{3J_H}{4}
  -\frac{J_t}{2}\right). \label{majocouplingsapp}
\end{eqnarray}

\section{Edge states}\label{edgestatesapp}
In this Appendix, we investigate the possible
existence of edge states in the non-degenerate
gapful phases Haldane charge, orbital and spin,
when open boundary conditions
are used. The nature of the edge states will
reinforce our characterization of, respectively,
the HC, HO, and RT phases as
being pseudo-spin 1 chains in, respectively, the
charge, orbital, and spin degrees of freedom.
Since the occurrence of edge states in the RT phase
has already been discussed at length in Ref.~\onlinecite{orignac},
we restrict our attention in the following to the HC and HO phases.

\subsection{Open boundary formalism}
The OBC are taken into account
by introducing two fictitious
sites $0$ and $N+1$ in Eq.~(\ref{GHundmodellattice}) and
by imposing vanishing boundary conditions on the fermion
operators: $c_0 = c_{N+1} = 0$~\cite{eggert,wong,gogo}.
The resulting boundary conditions on the Dirac fermionic fields
of Eq.~(\ref{contfer}) are thus:
\begin{eqnarray}
L_{l\sigma}\left(0\right) &=& - R_{l\sigma}\left(0\right) \nonumber \\
L_{l\sigma}\left(x = L\right) &=& -
\left(-1\right)^{L/a_0} R_{l\sigma}\left(x = L\right),
\label{boundfermxy}
\end{eqnarray}
with $L = (N+1)a_0$ and $l = 1,2$, $\sigma=\uparrow,\downarrow$.
The left and right-moving Dirac fermions are no longer independent
due to the presence of these open boundaries.
Using the bosonization formula (\ref{bosofer}), we deduce
the boundary conditions on the chiral bosonic fields:
\begin{eqnarray}
\Phi_{l\sigma L} \left(0 \right) &=& - \Phi_{l\sigma R} \left(0 \right)
+ \frac{\sqrt{\pi}}{2} + \sqrt{\pi} \; p_{l\sigma}
\nonumber \\
\Phi_{l\sigma L} \left(x = L \right) &=& - \Phi_{l\sigma R} \left(x = L \right)
+ \frac{\sqrt{\pi}}{2}\left(\frac{L}{a_0} - 1 \right)
\nonumber \\
 &+& \sqrt{\pi} \; q_{l\sigma} ,
\label{bcchiralboson}
\end{eqnarray}
where $p_{l\sigma}$ and $q_{l\sigma}$ are integers.
The total bosonic field $\Phi_{l\sigma}$ with internal
degrees of freedom $l\sigma$  thus obeys Dirichlet
boundary conditions:
\begin{eqnarray}
\Phi_{l\sigma} \left(0 \right) &=& \frac{\sqrt{\pi}}{2}
+ \sqrt{\pi} \; p_{l\sigma}
\nonumber \\
\Phi_{l\sigma} \left( x = L \right) &=& \frac{\sqrt{\pi}}{2}
\left( \frac{L}{a_0} - 1 \right)
+ \sqrt{\pi} \; q_{l\sigma} .
\label{dirichlet}
\end{eqnarray}
The next step of the approach is to introduce the mode decomposition
of the bosonic field $\Phi_{l\sigma}$ compatible with these
boundary conditions:
\begin{eqnarray}
&& \Phi_{l\sigma} \left(x, t\right) = \frac{\sqrt{\pi}}{2}
+ \frac{x}{L} \left[ \frac{\sqrt{\pi}}{2}
\left( \frac{L}{a_0} - 2 \right) + {\tilde \Pi}_{0 l\sigma}
\right]
\nonumber \\
&+& \sum_{n=1}^{\infty} \frac{\sin\left(n \pi x/L \right)}{\sqrt{n \pi}}
\left[ a_{n l\sigma} e^{- i n \pi v_F t/L}  + h. c.
\right],
\label{bosonmodedecomp}
\end{eqnarray}
where ${\tilde \Pi}_{0 l\sigma}$ is
the zero-mode operator with
spectrum $\sqrt{\pi} q_{l\sigma}$ and
$a_{n l\sigma}$ is the boson annihilation operator
obeying: $[a_{n l\sigma}, a^{\dagger}_{m l'\sigma'} ] =
\delta_{n,m} \delta_{l\sigma, l'\sigma'}$.
The mode decomposition of the dual field $\Theta_{l\sigma}$
can then be obtained from the property: $\partial_t \Theta_{l\sigma}
= v_F \partial_x \Phi_{l\sigma}$:
\begin{eqnarray}
&& \Theta_{l\sigma} \left(x, t\right) =  {\tilde \Phi}_{0 l\sigma}
+ \frac{v_F t}{L} \left[ \frac{\sqrt{\pi}}{2}
\left( \frac{L}{a_0} - 2 \right) + {\tilde \Pi}_{0 l\sigma}
\right]
\nonumber \\
&+i& \sum_{n=1}^{\infty} \frac{\cos\left(n \pi x/L \right)}{\sqrt{n \pi}}
\left[ a_{n l\sigma} e^{- i n \pi v_F t/L}  - h. c.
\right],
\label{dualbosonmodedecomp}
\end{eqnarray}
with $[{\tilde \Phi}_{0 l\sigma}, {\tilde \Pi}_{0 l'\sigma'} ] =
i \delta_{l\sigma,l'\sigma'}$.
In particular, $\Phi_{l\sigma}$ and $\Pi_{l\sigma} =
\partial_x \Theta_{l\sigma}$
satisfy the equal-time canonical commutation relation
for bosons:
$[\Phi_{l\sigma} (t,x), \Pi_{l'\sigma'} (t,y)] = i \delta_{l\sigma,l'\sigma'}
\delta_{L}(x-y)$, $\delta_{L}(x)$ being the delta function
at finite size: $\delta_{L}(x) = \sum_n e^{i n \pi x/L}/2L$.
Using the definitions $\Phi_{l\sigma R,L}
= (\Phi_{l\sigma} \pm \Theta_{l\sigma})/2$,
the mode decomposition of the chiral bosonic
fields $\Phi_{l\sigma R,L}$ can then be deduced
from Eqs.~(\ref{bosonmodedecomp},\ref{dualbosonmodedecomp}).
One can then show that these chiral fields satisfy the following
commutation relations when $L \gg a_0$:
\begin{eqnarray}
\left[\Phi_{l\sigma R,L} \left(x\right),
\Phi_{l'\sigma' R,L} \left(y\right)\right] &=& \mp \frac{i}{4}
\delta_{l\sigma, l'\sigma'} {\rm sgn} \left(x - y \right)
\label{commutOBC}\\
\left[\Phi_{l\sigma R} \left(x\right),
\Phi_{l'\sigma' L} \left(y\right)\right] &=& - \frac{i}{4}
\delta_{l\sigma, l'\sigma'} \; \;  0 < x,y < L,
\nonumber
\end{eqnarray}
${\rm sgn}(x)$ being the sign function.
At this point, one should note a technical subtlety which will
play its role in the investigation of the possible edge states
of the Haldane phases.
When considering OBC, the sign of the commutator
between $[\Phi_{l\sigma R} ,
\Phi_{l'\sigma' L}]$ is the opposite of the bulk one (\ref{commutator}).
The latter comes from the identity often used in
the bosonization approach~\cite{bookboso}:
\begin{equation}
\Phi_{l\sigma R,L} \left(x\right) = \frac{1}{2} \left(
\Phi_{l\sigma } \left(x\right) \mp \int_{- \infty}^{x}
dy \; \Pi_{l\sigma}\left(y\right) \right),
\label{techrmq}
\end{equation}
which does not take properly into account the boundary conditions
on the fields.
This subtlety has no effect on the derivation of the
low-energy Hamiltonian (\ref{Hamiltonian_tperp0_majorana}) which is still
valid in presence of OBC as it can be easily shown.
However,  it will be important for the
discussion of edge states as first observed in
Ref.~\onlinecite{orignac} for the
determination of boundary excitations
of the semi-infinite two-leg spin ladder.

With this formalism at hands, we are now in position to investigate
the possible edge states in the generalized Hund model
(\ref{GHundmodellattice}) with OBC. To simplify the discussion,
we will consider a semi-infinite geometry where
the OBC is located at the $i=0$ site and  $L \rightarrow
+ \infty$.
The low-energy effective Hamiltonian density is still
given by Eq.~(\ref{Hamiltonian_tperp0_majorana}),
but now we see, using the bosonic boundary conditions
(\ref{bcchiralboson}), the change of basis (\ref{basebosons}),
and the refermionization (\ref{refer}),
that the eight Majorana fermions $\xi^A_{R,L}$ ($A = 1, \ldots, 8$)
must verify the following boundary conditions:
\begin{equation}
\xi^A_{L} \left(0\right) = \xi^A_{R} \left(0\right) .
\label{bcMajorana}
\end{equation}

\subsection{Edge states in the Haldane charge phase}

Let us set our investigation on the symmetry line $J_H=-8U/3$ where
the charge degrees of freedom display an extended $SU(2)$ symmetry and
form the pseudo-spin 1 in charge.  For strong attractive $U$, we
expect that the spin and orbital gaps will be larger than the charge
gap.  This is confirmed numerically by DMRG calculations: for
instance, for $U/t=-3$ and $J_t=0$, the spin and orbital gaps are
close respectively to $6t$ and $5t$, while the charge gap is roughly
$0.1t$.  Within this hypothesis, we can safely integrate out spin and
orbital degrees of freedom of model
(\ref{Hamiltonian_tperp0_majorana}). For simplicity, let us choose to
work on the line $J_t=J_H/4$, where the spin and orbital degrees of
freedom are unified. Keeping only the relevant terms and again
neglecting velocity anisotropy, the resulting effective Hamiltonian
then reads:
\begin{eqnarray}
  {\cal H} &=&
- \frac{i v_c}{2} \sum_{a=6}^{8} \int_{0}^{\infty} d x \; \left(
\xi_R^{a} \partial_x \xi_R^{a}
- \xi_L^{a} \partial_x \xi_L^{a} \right)
\nonumber \\
&-& i m_c \int_{0}^{\infty} d x \;
\sum_{a=6}^{8} \xi_R^{a} \xi_L^{a} ,
\label{majochargeOBC}
\end{eqnarray}
with the mass $m_c$ given by ($g_7=g_3$
on the line $J_H=-8U/3$):
\begin{eqnarray}
m_c &=&  i g_7 \sum_{a=1}^{5}
\langle \xi_R^{a} \xi_L^{a}
\rangle .
\label{massescharge}
\end{eqnarray}

Model (\ref{majochargeOBC}) is the sum of three decoupled semi-infinite
free massive Majorana fermion models. Hence, let us now consider a
single Majorana fermion $\xi_{R,L}$ model:
\begin{equation}
{\cal H}_m =
- \frac{i v}{2}  \int_{0}^{\infty} d x \; \left(
\xi_R \partial_x \xi_R
- \xi_L \partial_x \xi_L \right)
- i m \int_{0}^{\infty} d x
\xi_R \xi_L ,
\label{toyMajomodel}
\end{equation}
with boundary condition: $\xi_L (0) = \xi_R (0)$.
Model (\ref{toyMajomodel}) is quadratic
with dispersion relation $\epsilon_k = \sqrt{v^2 k^2 + m^2}$
and with the fermionic decomposition~\cite{orignac}:
\begin{eqnarray}
\xi_R \left(x,t\right) &=& \frac{1}{\sqrt{2L}}
\sum_{k>0} \left\{ \xi_k \left(\cos
\left(kx+\theta_k\right) + i \sin\left(kx\right) \right)
e^{-i \epsilon_k t}  \right.
\nonumber \\
 &+&  \left. h. c.  \right\} +
\sqrt{\frac{m}{v}} \; \theta\left(m\right) e^{-m x/v} \eta
\nonumber \\
\xi_L \left(x,t\right) &=& \frac{1}{\sqrt{2L}}
\sum_{k>0} \left\{ \xi_k \left(\cos
\left(kx+\theta_k\right) - i \sin\left(kx\right) \right)
e^{-i \epsilon_k t}  \right.
\nonumber \\
 &+&  \left. \hc  \right\} +
\sqrt{\frac{m}{v}} \; \theta\left(m\right) e^{-m x/v} \eta
,
\label{decompmajo}
\end{eqnarray}
where $\xi_k$ is fermion annihilation operator with wave-vector $k$,
$\theta$ is the step function, and $\eta$ is a zero mode
real fermion normalized according
to $\eta^2 = \frac{1}{2}$. In Eq.~(\ref{decompmajo}), $\theta_k$ is
defined by:
\begin{equation}
\exp \left(i \theta_k \right) = \frac{ v k + i m}{\epsilon_k}.
\end{equation}
The key point of Eq.~(\ref{decompmajo}) is the existence
of an exponentially localized Majorana state with zero
energy inside the gap (midgap state) for a positive mass $m$.
In contrast, for negative $m$, such a zero mode contribution
does not occur since it is not a normalizable solution.

The presence of edge states for the HC phase
thus depends on the sign of the mass $m_c$. Using the
definition (\ref{refer}) and the commutator
(\ref{commutOBC}), we find:
\begin{eqnarray}
m_c &=&  - \frac{g_7}{\pi a_0} \langle C^{\sqrt{4 \pi}}_{s} +
C^{\sqrt{4 \pi}}_{f} \nonumber\\
&&\qquad+ \frac{1}{2}
\left( C^{\sqrt{4 \pi}}_{sf} +
{\tilde C}^{\sqrt{4 \pi}}_{sf} \right) \rangle .
\label{masseschargebose}
\end{eqnarray}

In the HC phase, we have: $\langle\Phi_s
\rangle=\langle\Phi_f \rangle=\langle\Theta_{sf} \rangle = 0$
(see Table II),
so that $m_c=-5g_7/2\pi$. Hence, since $g_7=a_0 U$,
the mass $m_c>0$, which signals the emergence of
three localized Majorana modes $\eta^{a}$ ($a=6,7,8$)
from the mode decomposition (\ref{decompmajo}).
Moreover, three local Majorana fermion modes are known to
define to a local pseudo spin-$\frac{1}{2}$ operator ${\cal J}^a$ thanks
to the identity~\cite{tsvelikmajo}:
\begin{equation}
{\cal J}^{a} = - \frac{i}{2} \epsilon^{a b c} \eta^{5+b} \eta^{5+c},
\label{spin1demiedgestate}
\end{equation}
that is a consequence of the anticommutation relations $\{\eta^{a} , \eta^{b} \} =
\delta^{ab}$. We thus conclude on the existence, in the HC phase, of a pseudo
spin-$\frac{1}{2}$ edge
state at the boundary which can be viewed as a holon edge state.

One recognizes that the pseudo-spin projection along ${\cal J}^1=-i\eta^7\eta^8$
is proportional to the total charge generating U(1)$_c$: in the continuum limit,
within the convention (\ref{refer}),
one has~\cite{orignac} $Q_c=\frac{1}{2}\sum_i n_i \longrightarrow
-i\int dx \left(\xi^7_R\xi^8_R + \xi^7_L\xi^8_L\right) =
{\cal J}^1 -2i\sum_{k>0} \left( \xi^7_k\xi^{8\dagger}_k - \xi^8_k\xi^{7\dagger}_k \right) $,
showing that the zero-mode contributes the total charge. In a finite size system of size $L$
with two boundaries, the edge states come into pairs, that organize into a pseudo-spin
singlet and a pseudo-spin triplet. Edge states at the two end of the chain interact,
leading to a singlet/triplet splitting that goes to zero in the thermodynamical limit.
It results that one observes a mid-gap state with quantum numbers $(Q_c,S^z,T^z)=(L\pm 1,0,0)$.

\subsection{Edge states in the Haldane orbital phase}
Let us now sit on the symmetry line $J_t=-3J_H/4$ where it is the
orbital degrees of freedom that display an extended SU(2) symmetry and
form a pseudo-spin 1. For the sake of simplicity, let us
choose to look at line $J_H=8U$ when
charge and spin degrees of freedom are unified into an SO(5)
symmetry. For repulsive $U$, we expect that the charge and spin gap
will be higher than the orbital gap (which is confirmed numerically by
DMRG simulations for instance for $U/t=1$: the charge and spin gaps
are both equal to $6.4t$ while the orbital gap is $0.08t$) so that we
can safely integrate out the corresponding degrees of freedom. The
resulting leading effective Hamiltonian is (neglecting velocity
anisotropy):
\begin{eqnarray}
  {\cal H} &=&
- \frac{i v_t}{2} \sum_{a=4}^{6} \int_{0}^{\infty} d x \; \left(
\xi_R^{a} \partial_x \xi_R^{a}
- \xi_L^{a} \partial_x \xi_L^{a} \right)
\nonumber \\
&-& i m_o \int_{0}^{\infty} d x \;
\sum_{a=4}^{6} \xi_R^{a} \xi_L^{a} ,
\label{majoorbitalOBC}
\end{eqnarray}
with the mass $m_o$ given by:
\begin{eqnarray}
m_o &=&  i g_8 \sum_{a=1}^{5}
\langle \xi_R^{a} \xi_L^{a}
\rangle .
\label{massesorbital}
\end{eqnarray}

The latter mass can be expressed in terms of the
ground state expectation value of the
bosonic fields for the charge and spin degrees of freedom:
\begin{eqnarray}
m_o &=&  - \frac{g_8}{\pi a_0} \langle C^{\sqrt{4 \pi}}_{c} +
C^{\sqrt{4 \pi}}_{s} \nonumber\\
&&\qquad+ \frac{1}{2}
\left( C^{\sqrt{4 \pi}}_{sf} +
{\tilde C}^{\sqrt{4 \pi}}_{sf} \right) \rangle .
\label{massesorbitalbose}
\end{eqnarray}
so that $m_c=-5g_8/2\pi$ in the HO phase (see Table II). As on the considered line,
$g_8=-2a_0 U$,
the mass $m_o>0$. We thus conclude on the existence
of three localized Majorana modes which form a pseudo
spin-$\frac{1}{2}$ (orbital) edge state at the boundary in the
HO phase.

Repeating the same argument as in the HC phase leads to the expectation
-- in a finite geometry with size $L$ and two OBC  --
of a mid-gap state with quantum numbers $(Q_c,S^z,T^z)=(L,0,\pm 1)$.

\section{Strong coupling around the orbital line}
\label{appendixTperp}

In this Appendix we give a description of the effect of an interchain hopping $t_\perp$
in the strong coupling regime. We will first show that
close to the orbital symmetric line ($J_t=-3J_H/4$ ),
a low-energy effective continuous theory can be derived and
trivially solved, allowing for the
identification of the phases and the phase transitions. Then, we will show
that a $\widetilde{\mbox{U(1)}}$
symmetry emerges at low energy, similarly to what is known
at weak coupling (see Section III D).

\subsection{Continuous theory}

The effect of the inter-leg hopping in the strong coupling regime is in
general a complicated problem: since
the hopping term breaks the U(1) orbital symmetry,
this process will induce transitions amongst on-site states
(depicted in Fig. \ref{energylevels})
that belong to different symmetry multiplets.

However, as noticed in Section \ref{strongcoupling}, for a  special fine-tuning
 of the couplings $J_t=-3J_H/4$ the lattice
model enjoys a SU(2) symmetry in the orbital sector. Close to this line,
the orbital symmetric line,
a strong coupling expansion can be performed: orbital degrees
of freedom are the only low-energy modes, and  an effective
Hamiltonian for the orbital operators $T^a$ can be derived,
that governs their dynamics. Noticing that the interchain
hoping term can be expressed in terms of orbital degrees of freedom,
$t_\perp
\sum_{\sigma,i}(c^\dagger_{1\sigma,i}c_{2\sigma,i}\dagg
+c^\dagger_{2\sigma,i}c_{1\sigma,i}\dagg) = 2t_\perp \sum_i T^x_i$,
one remarks that the effect of $t_\perp$ close to
the SU(2)$_o$ line amounts to the analog of a transverse magnetic
field in direction $x$ for orbital degrees of freedom,
resulting in the following effective Hamiltonian:
\begin{equation}
 \mathcal{H}_{\text{eff}}=J_o  \sum_i \vec{T}_i \cdot \vec{T}_{i+1} +
 D_o \sum_i(T^z_i)^2 + h_o \sum_i T_i^x,
\label{effectiveTx}
\end{equation}
where $J_o=16t^2 /(9J_H+ 8U )$, $D_o=J_t+3J_H/4$ and
$h_o=2t_\perp$. Orbital operators $T_i$ being spin-one operators, close
to the orbital line the problem is thus equivalent to
 a spin-1 Heisenberg model with single-ion anisotropy
under a transverse magnetic field.

In the absence of a magnetic field $h_o$, it is known that the spin-1
Heisenberg chain
with a single-ion anisotropy can be described in terms
of continuous degrees of freedom, namely 3 Majorana fermions~\cite{tsvelikspin1} (to be consistent with the main text, we call
them $\xi^a$, $a=4,5,6$), that are related as follows to
the uniform components of the spin operators:
$T^a_i \to T^a(x)= i\frac{\epsilon^{abc}}{2}\left(\xi^{3+b}_L\xi^{3+c}_L
+ \xi^{3+b}_R\xi^{3+c}_R\right) + \ldots$,
where the dots indicate oscillating terms. The low energy spectrum of
the theory is well described, at lowest order, by a theory of 3 free
massive fermions, corresponding to 3 branches of magnons.
Now the magnetic field also yields a term
that is quadratic in fermions. Neglecting velocity anisotropies,
we thus end up with the following quadratic Hamiltonian:
\begin{eqnarray}
\mathcal{H}_{\text{cont}}&=&
-\frac{i v_o}{2} \sum_{a=4}^6 (\xi_R^a \partial_x
  \xi_R^a - \xi_L^a \partial_x \xi_L^a) \nonumber\\
&&  +\; i\sum_a m_a\xi^a_R\xi^a_L + ih_o \left(\xi^{5}_L\xi^{6}_L + \xi^{5}_R\xi^{6}_R\right).
\end{eqnarray}
The masses  $m_a$ are phenomenological parameters.
 When $D_o=t_\perp=0$, one has a single mass scale $m\propto J_o$ (the gap of
 the spin-1 Heisenberg chain),
 and in the general case one can parametrize them as
 $m_4=m_5=m-\delta$ and $m_6=m+\delta$, with
 $\delta\propto D_o$ at first order in $D_o/J_o$.
 One immediately sees that the fermion $\xi^4$ decouples.
 Fourier transforming the remaining fermions,
 $\xi_{L(R)}^a(k)=\int dx\, e^{-ikx}\xi^a_{L(R)}(x)$,
 and introducing $\boldsymbol{\Psi}_k=(\xi_L^5(k),\xi_R^5(k),\xi_L^6(k),\xi_R^6(k))$
 the Hamiltonian reads
 $\sum_{k>0} \boldsymbol{\Psi}_{-k}U_k\boldsymbol{\Psi}_k$.
The one-particle spectrum is obtained from
 the eigenvalues of the matrix:
\begin{equation}
U_k = \left(\begin{array}{cccc}
v_o k & im_5 & ih_o & 0 \\
-im_5 & - v_o k &0& ih_o\\
- ih_o & 0 & v_o k & im_6\\
0& -ih_o & -im_6 & - v_o k
\end{array}   \right) .
\end{equation}
This yields two branches with energies:
$\epsilon_\pm(k) = \sqrt{\epsilon_0^2(k) + h_0^2 + \delta^2 \pm 2\sqrt{h_o^2\epsilon_0^2(k)+m^2\delta^2}}$, where $\epsilon_0(k)$ is the spectrum at
$D_o=t_\perp=0$: $\epsilon_0(k)=\sqrt{v_o^2k^2 + m^2}$.

We can now identify several phase transition lines, for
which the spectrum is massless, i.e., admits modes of
arbitrary low energy. A first Ising transition line
is readily obtained when $\delta=m$: then $m_4=0$.

Other critical lines are found by solving the equation
$\epsilon_-(k^*)=0$ (the branch $\epsilon_+(k)$ is always gapful):
one finds that for $h_o^2+\delta^2=m^2$, this equation has a solution
$k^*=0$, and for $\delta=0$ and $|h_o|>m$, two solutions
$k^*=\pm|k^*|\neq 0$, with $v_o^2k^{*2}=h_o^2-m^2$. On the former critical line,
the massless degrees of freedom consist in a single
Majorana mode and one has a central charge  $c=\frac{1}{2}$.
The latter critical line has central charge $c=1$,
and corresponds to the well-known commensurate/incommensurate
transition of the isotropic spin-1 Heisenberg chain under a magnetic field.

\begin{figure}[h]
\includegraphics[width=0.9\linewidth]{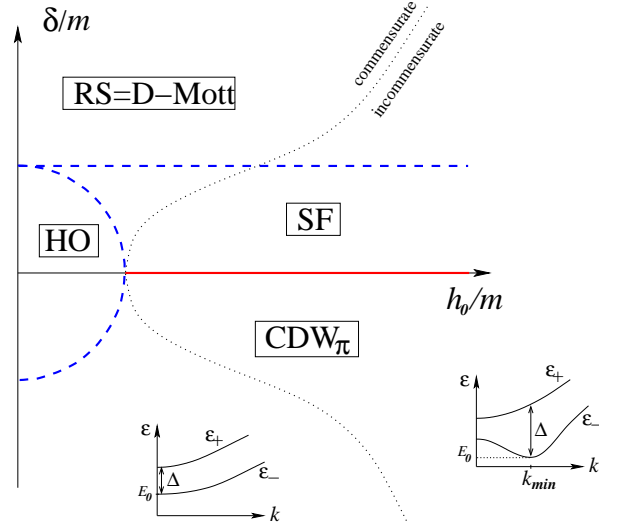}
\caption{(Color online) Phase diagram of the generalized Hund ladder with
interchain hopping. The coordinate
on the horizontal axis
is proportional to $t_\perp$ whereas that on the vertical axis is the distance
$D_o$ to the SU(2) orbital symmetric model. Blue lines indicate
$c=\frac{1}{2}$ critical lines, and the red one is a $c=1$ critical line.
See the main text for the
definition of the phases. The dashed line, with equation
$m^2\delta^2=h_o^2(h_o^2-m^2)$, indicates the location
where incommensuration appears: on the right of this curve,
the lower band $\epsilon_-$ displays a minimum for a wave
vector $k_{min}$. The two insets display the typical spectrum
in the commensurate and incommensurate regions, respectively.}
\label{phaseStrongOrbital}
 \end{figure}

To identify the phases of the ladder lying on both sides of those lines
(see Fig. \ref{phaseStrongOrbital}),
one notices that by continuity with what happens at
$t_\perp=0$, the phase with $m_4<0$
 (obtained at large enough $\delta$) must be a
 RS phase (which coincides with the D-Mott phase).
 The phase at $m_4>0$ is readily identified by means
 of the duality transformation $\xi^4_L\to-\xi^4_L$ that
 amounts to $\Phi_f\leftrightarrow\sqrt{\pi}/2 - \Theta_f$
(see Eqs.~(\ref{refer})): it is a SF phase.

 By continuity with $t_\perp=0$, the pocket
 $\delta^2+h_o^2<m^2$ corresponds
to a HO phase. The last phase to be mapped
out is the one lying at $h_o^2+\delta^2>m^2$ and $\delta<0$: it is adiabatically
connected to the ODW phase that one has obtained at $t_\perp=0$ and $D_o<0$ so
that it is an ODW or CDW$_\pi$ phase, as we named it in the context of the ladder
with an interchain hopping.

\subsection{Effective low-energy theory}

We now make a qualitative connection with the weak-coupling analysis of
Section \ref{subsectionTperp}, which crucially relies on the existence of
an extended  $\widetilde{U(1)}$ symmetry in the orbital
sector, and show that this symmetry is also emergent at
low energy in the strong-coupling regime.
In the strong-coupling limit, in the vicinity of the orbital symmetric line,
the interchain hopping $t_\perp$ amounts to a magnetic field in orbital
space, that will eventually drive the system to a
incommensurate state: when $\delta=0$,
as soon as $h_o>m$ the Fermi points
shift to $k_F=\frac{\pi}{2a_0}+k^*$. Now if the ``magnetic field''
is large enough, i.e., if $k^*$ is large enough, this leaves the space
for a low-energy
description built on fields expanded around the new Fermi points.
In this effective low-energy
description, $4k_F$ umklapp terms have no effect as being strongly oscillating.
This remains
true if one departs from the orbital symmetric
line $\delta=0$: in this case, the spectrum of the
low energy band, $\epsilon_-(k)$, develops a gap but the
incommensuration is still present in the
form of a minimum (which becomes infinitely deep
in the limit $h_o\gg m$) in $\epsilon_-(k)$
at a value $v_ok_{min}=\sqrt{h_o^2-m^2(1+\frac{\delta^2}{h_o^2})}$.
This happens (see Fig. \ref{phaseStrongOrbital}) when $\delta$ is small enough,
$|\delta|<\delta_c=h_o\sqrt{\frac{h_o^2}{m^2}-1}$.
But is there an associated
emergent symmetry, as it was the case in the weak-coupling limit?

To investigate this, one has to enter into further details.
Denoting by $A_k$ the unitary
matrix diagonalizing $U_k$, with
$A_k^\dagger U_k A_k=\mbox{Diag}(\epsilon_-(k),-\epsilon_-(k),\epsilon_+(k),-\epsilon_+(k))$,
and introducing
$\widetilde{\boldsymbol{\Psi}}_k=A_k^{-1}\boldsymbol{\Psi}_k=(\chi_-(k),
\chi_+(k))$,
one can represent in the eigenmode basis the orbital U(1) generators along $x$,
$T^x_{tot,\pm}=i\int \,dx\,\left(\xi^5_L\xi^6_L\pm\xi^5_R\xi^6_R\right)
= \int_{k>0}\widetilde{\boldsymbol{\Psi}}_k^\dagger T^{(\pm)}_k
\widetilde{\boldsymbol{\Psi}}_k$. $T^x_{tot,\pm}$
are the total Noether charge and current. The Hermitian matrix elements
 $T^{(\pm)}_{k,\alpha\beta}$ are  too cumbersome, and not particularly enlightening,
 to appear here, but the matrix $T^{(-)}_{k}$ has a remarkable block structure:
\begin{equation}
T^{(-)}_k=\left(
\begin{array}{cc}
t_1(k)\mathbb{I} & t_2(k)\sigma^z+t_3(k)\sigma^x\\
 t_2(k)\sigma^z+t_3(k)\sigma^x & -t_1(k)\mathbb{I}
\end{array}
\right).
\end{equation}

Let us now introduce two quantities characterizing the spectrum: the absolute
 minimum $E_0$ of the lower band, and the gap $\Delta$ from the minimum
 of the lower band $E_0$ to the upper band (see insets of Fig.\ref{phaseStrongOrbital}).
Now, if the two bands $\epsilon_+(k)$ and $\epsilon_-(k)$ are
well separated, i.e. if  $\Delta\gg E_0$,
it makes sense to describe the theory at low energies in terms
of the two Majorana fermions $\chi_-(k)$.
Introducing the projector $P^-$ that projects on this subspace,
 the effective Hamiltonian
$H_{\rm eff}^- = P^- H P^- = \sum_k\epsilon_-(k)\chi_-^\dagger(k)\sigma_z\chi_-(k)$
 \emph{commutes} with
$P^-\; T^x_{tot,-}\;P^-$. Thus, the total orbital current
along $x$ is asymptotically conserved
in the limit of large $\Delta/E_0$, and the low-energy theory
is effectively $\widetilde{\mbox{U(1)}}$
symmetric. For a physical quantity computed in the low-energy,
$\widetilde{\mbox{U(1)}}$-symmetric theory, violation of this emergent
symmetry by processes connecting
the two bands will result in corrections of order $(E_0/\Delta)^2$.

Now, the parameters $E_0$ and $\Delta$ bear qualitatively distinct
forms in the commensurate and incommensurate regions.
For simplicity, we restrict our attention to the case $h_o>m$.
In the commensurate region, one has:
\begin{eqnarray}
E_0&=&\sqrt{\delta^2+h_o^2}-m,\nonumber\\
\Delta &=& 2m,
\end{eqnarray}
so that in general $E_0/\Delta$ is not small.

In contrast, in the incommensurate region,
one has:
\begin{eqnarray}
E_0 &=& \delta\sqrt{1-\frac{m^2}{h_o^2}} \;\lesssim\; {\cal O}(\delta), \nonumber\\
\Delta &=& \sqrt{4h_o^2+E_0^2}-E_0,
\end{eqnarray}
so that the symmetry  $\widetilde{\mbox{U(1)}}$ is violated by terms of
order $(\delta/2h_o)^2$: the symmetry $\widetilde{\mbox{U(1)}}$ is asymptotically
exact in the limit $h_o\gg\delta$. One thus concludes that close to the line
$J_t=-3J_H/4$, the situation at large coupling coincides with that at small coupling.
One can thus reasonably infer that
this symmetry is emergent in all regimes at
least close to the line $J_t=-3J_H/4$. Our strong coupling analysis suggests
that the emergent $\widetilde{\mbox{U(1)}}$ is broken when moving
far away from the orbital symmetric line ; however, the question of the existence of such
an emergent symmetry in the strong-coupling regime of
the generic electronic two-leg ladder goes far beyond the scope of this paper.


\begin{thebibliography}{99}
\bibitem{comment}
This phase is equivalent to the ODW phase for $t_{\perp} = 0$.
\bibitem{comment2}
This phase is equivalent to the staggered dimerization
phase (SP$_\pi$ phase) that we have found for $t_{\perp} = 0$.
\end{thebibliography}
\end{document}